\pdfoutput=1
\documentclass[11pt,aps,prd,a4paper,showpacs,showkeys,superscriptaddress, preprintnumbers,floatfix,nofootinbib,notitlepage]{revtex4-1}

\usepackage[toc,page]{appendix}
\usepackage{latexsym}
\usepackage{revsymb}
\usepackage{multirow}
\usepackage{color}
\usepackage[usenames,dvipsnames,svgnames,table]{xcolor}

\usepackage{graphicx}
\usepackage{epsfig}  
\usepackage{epsf}    
\usepackage{dcolumn}
\usepackage{bm}
\usepackage{textcomp}
\usepackage{float}
\usepackage[]{hyperref}
\usepackage{adjustbox}
\usepackage[utf8]{inputenc}
\usepackage[T1]{fontenc} 
\usepackage{url}
\usepackage{amsmath}
\usepackage{mathtools}
\usepackage{subfigure}
\usepackage{alphalph}
\usepackage{morefloats}
\usepackage{comment}
\usepackage{ulem}

\hypersetup{colorlinks=true,linkcolor=blue,citecolor=blue,urlcolor=blue}

\newcommand{\AddrIPN}{%
Departamento de F\'{\i}sica, Centro de Investigaci\'on
  y de Estudios Avanzados del IPN,\\ Apartado Postal 14-740 07000,
  Ciudad de Mexico, Mexico}
  
\newcommand{\AddrIPhT}{Institut de Physique Th{\'e}orique, Universit{\'e} Paris Saclay, CNRS, CEA, F-91191 Gif-sur-Yvette, France}  

\newcommand{\AddrKIT}{Institut f\"{u}r Astroteilchenphysik, Karlsruher Institut f\"{u}r Technologie (KIT), 76021 Karlsruhe, Germany}  

\newcommand{\AddrIFIC}{Instituto de F\'isica Corpuscular (CSIC-Universitat de Val\`encia),\\Parc Cient\'ific UV, C/ Catedr\`atico Jos\'e Beltr\'an, 2, 46980 Paterna, Spain}

\begin{document}

\title{Exploring the Sensitivity to Non-Standard Neutrino Interactions \\ of NaI and Cryogenic CsI Detectors at the Spallation Neutron Source}

\author{Sabya Sachi Chatterjee} \email{sabya.chatterjee@kit.edu} \affiliation{\AddrKIT}\affiliation{\AddrIPhT}

\author{St{\'e}phane Lavignac} \email{stephane.lavignac@ipht.fr} \affiliation{\AddrIPhT}

\author{O. G. Miranda} \email{omar.miranda@cinvestav.mx} \affiliation{\AddrIPN}

\author{G. Sanchez Garcia} \email{gsanchez@ific.uv.es} \affiliation{\AddrIFIC}

\begin{abstract}
After the first observation of coherent elastic neutrino-nucleus scattering (CE$\nu$NS) by the COHERENT collaboration,
many efforts are being made to improve the measurement of this process,
making it possible to constrain new physics in the neutrino sector. 
In this paper, we study the sensitivity to non-standard interactions (NSIs) and generalized neutrino interactions (GNIs) of two experimental setups at the Spallation Neutron Source at Oak Ridge National Laboratory: a NaI detector with characteristics similar to the one that is currently being deployed there, and a cryogenic CsI detector proposed at the same facility.
We show that a combined analysis of the data from these detectors, whose target nuclei
have significantly different proton-to-neutron ratios,
could help to partially break the parameter degeneracies arising from the interference between the Standard Model
and NSI  contributions to the CE$\nu$NS cross section,
as well as between different NSI parameters.
By contrast, only a slight improvement over the current CsI constraints is expected
for parameters that do not interfere with the SM contribution. 
\end{abstract}

\maketitle

\section{Introduction}
\label{sec:introduction}

Coherent elastic neutrino-nucleus scattering (CE$\nu$NS)~\cite{PhysRevD.9.1389} is a privileged process
to probe new physics in the neutrino sector.
So far, the only measurements of CE$\nu$NS have been done at the Spallation Neutron Source (SNS)
at Oak Ridge National Laboratory, USA, by the COHERENT collaboration.
The first observation of this process was performed with a CsI detector~\cite{Akimov:2017ade,Akimov:2021dab}.
Another detector, with liquid argon as a target, was subsequently used by the COHERENT collaboration~\cite{COHERENT:2020iec}.
These measurements allowed to test the Standard Model at low energy~\cite{Cadeddu:2021ijh},
to constrain the neutron root mean square {\it (rms)} radius of cesium, iodine~\cite{Cadeddu:2017etk}
and argon~\cite{Miranda:2020tif}, and to probe various possible manifestations of physics beyond the Standard Model,
such as non-standard neutrino interactions (NSIs)~\cite{Barranco:2005yy, Scholberg:2005qs, DeRomeri:2022twg,Breso-Pla:2023tnz},
generalized neutrino interactions (GNIs)~\cite{Altmannshofer:2018xyo,AristizabalSierra:2018eqm,Flores:2020lji, Han:2020pff, DeRomeri:2022twg},
neutrino electromagnetic properties~\cite{Cadeddu:2019eta,Miranda:2019wdy,Cadeddu:2020lky},
new light gauge bosons~\cite{Farzan:2018gtr,Flores:2020lji,delaVega:2021wpx,Bertuzzo:2021opb,Denton:2022nol}
and sterile neutrinos~\cite{Dutta:2015nlo,Kosmas:2017zbh,DeRomeri:2022twg}, among others. Future CE$\nu$NS experiments
at spallation neutron sources~\cite{Akimov:2022oyb, Baxter:2019mcx, CCM:2021leg}
or nuclear reactors~\cite{CONUS:2021dwh, CONNIE:2016nav, nGeN:2022uje, akimov2022red}
will provide more data using different target materials. Particularly interesting is the possibility to disentangle
new physics from Standard Model parameters using the interplay between experiments with different neutrino
sources~\cite{Canas:2019fjw, Rossi:2023brv}. In addition, using nuclear targets with different proportions of protons and neutrons
can help to reduce the parameter degeneracies arising from the interference between standard
and non-standard contributions to CE$\nu$NS, e.g., in the presence of NSIs~\cite{Barranco:2005yy}.

A broad class of new physics models predict low-energy effects in the lepton sector
that can effectively be described in terms of NSIs or GNIs~\cite{Lindner:2016wff, Ohlsson:2012kf,Miranda:2015dra,Farzan:2017xzy}.
CE$\nu$NS is an excellent process to test such new physics effects and to resolve the degeneracies
that appear in the interpretation of neutrino oscillation data in the presence of NSIs~\cite{Miranda:2004nb,Coloma:2016gei}.
However, the possibility of constraining NSI parameters with CE$\nu$NS is limited
by the possible cancellations between standard and non-standard contributions
to the cross section, as well as between different NSI couplings. It is therefore crucial
to perform CE$\nu$NS measurements with different neutrino sources and targets
in order to break  these degeneracies~\cite{Barranco:2005yy, Scholberg:2005qs, Baxter:2019mcx, Galindo-Uribarri:2020huw, Coloma:2016gei,Chatterjee:2022mmu}.

In this work, we study the expected sensitivities to non-standard and generalized neutrino interactions
of two experimental setups at the Spallation Neutron Source: a NaI detector with characteristics similar
to the one that is currently being deployed there~\cite{Hedges:2021mra},
expected to be sensitive only to sodium recoils~\cite{Cintas:2024pdu},
and a cryogenic CsI detector proposed at the same facility~\cite{COHERENT:2023sol}.
The target nuclei of these detectors have significantly different proton-to-neutron ratios
($p/n \simeq 0.71, 0.72$ for cesium and iodine, and $p/n \simeq 0.92$ for sodium).
As we will show, this important feature makes a combined analysis of the data from both detectors
particularly efficient at breaking degeneracies involving flavor-diagonal NSI parameters.
For generalized neutrino interactions and flavor off-diagonal NSIs,
whose contribution to the CE$\nu$NS cross section does not interfere with the SM one,
no significant improvement over the constraints extracted from
the current CsI data is expected. 

The paper is structured as follows. In Section~\ref{sec:theory}, we discuss the NSI and GNI contributions
to the CE$\nu$NS cross section. In Section~\ref{sec:experiment}, we present the experimental setup
at the SNS and the characteristics of the current CsI and future NaI and cryogenic  CsI detectors. 
Section~\ref{sec:analysis} details our analysis procedure.
The expected sensivities of the future detectors, and of their combination,
to NSI and GNI parameters are presented in Section~\ref{sec:results}, and compared with the current CsI constraints.
Finally, we present our conclusions in Section~\ref{sec:conclusions}.

\section{Theoretical framework}
\label{sec:theory}

In this section, we discuss the necessary tools to probe GNI through CE$\nu$NS measurements. We begin by introducing the SM prediction for the CE$\nu$NS cross section and then we discuss how this cross section is modified when GNI parameters are introduced. 
 
\subsection{The standard CE$\nu$NS cross section}
\label{subsec:CEvNS_SM}

In the Standard Model, the cross section for the coherent elastic scattering
of a neutrino or antineutrino of flavor $\alpha$ ($\alpha = e, \mu, \tau$) and energy $E_\nu$ off a nucleus
with $Z$ protons, $N$ neutrons and mass $M$ is given (up to subleading terms of order $T/E_\nu$,
$(T/E_\nu)^2$~\cite{Barranco:2005yy} and small radiative corrections~\cite{Tomalak:2020zfh}) by~\cite{PhysRevD.9.1389} 
\begin{eqnarray}
\frac{\mathrm{d}\sigma}{\mathrm{d}T} (E_\nu, T)\, =\, \frac{G_F^2M}{\pi}\left(1-\frac{MT}{2E^2_\nu}\right)
  F^2(|\vec{q}|^2)\, Q_{W,\alpha}^2\ ,
\label{eq:CEvNS_SM}
\end{eqnarray}
where $G_F$ is the Fermi constant,
$T$ the nuclear recoil energy, $F(|\vec{q}|^2)$ the form factor of the nucleon
distribution in the nucleus~\cite{Lewin:1995rx} (assumed to be the same for protons and neutrons)
evaluated at the transferred three-momentum $|\vec{q}| \simeq \sqrt{2MT}$, and
\begin{eqnarray}
Q_{W,\alpha}\, =\, Z g_V^p + N g_V^n\, ,
\label{eq:weak_charge}
\end{eqnarray}
with $g_V^p = \frac{1}{2}-2\sin^2\theta_W$ and $g_V^n = -\frac{1}{2}$, is the weak nuclear charge.
Since $|\vec{q}|$ is typically very small in CE$\nu$NS, the value of the weak mixing angle
is taken at zero momentum transfer ($\sin^2\theta_W = 0.23867$~\cite{Erler:2017knj,ParticleDataGroup:2022pth}).

The form factor $F(|\vec{q}|^2)$ describe the loss of coherence of the scattering when $|\vec{q}| \gtrsim R^{-1}$,
where $R$ is the nucleon radius.
Following the COHERENT collaboration, we use the Klein-Nystrand parametrization
for the nuclear form factor~\cite{PhysRevC.60.014903},
\begin{eqnarray}
F(|\vec{q}|^2)\, =\, 3\, \frac{j_1(|\vec{q}|R_A)}{|\vec{q}|R_A} \left(\frac{1}{1+|\vec{q}|^2a_k^2} \right)\, ,
\end{eqnarray}
where $j_1 (x) = (\sin x - x \cos x)/x^2$ is the spherical Bessel function of order one,
$a_k = 0.9\, {\rm fm}$ and $R_A = 1.23A^{1/3}\, \mbox{fm}$ is the nuclear radius,
with $A$ the total number of nucleons of the target nucleus.

\subsection{Non-standard and generalized neutrino interactions}
\label{subsec:GNIs}

The CE$\nu$NS cross section is potentially affected by any interaction between neutrinos
and quarks that may be generated by new physics beyond the Standard Model.
The simplest example of such interactions are the so-called (neutral-current) non-standard neutrino
interactions (NC-NSIs, or NSIs for short), whose phenomenology has been widely studied (see e.g. the reviews~\cite{Ohlsson:2012kf,Miranda:2015dra,Farzan:2017xzy, Giunti:2019xpr} and references therein).
They are usually parametrized as
\begin{equation}
\mathcal{L}_{\mathrm{NC\mbox{-}NSI}}\, = \,
- \sqrt{2}\, G_F
\left\{
\varepsilon_{\alpha\beta}^{fV}\,
\bigl(\overline{\nu}_\alpha\gamma^\mu P_L \nu_\beta\bigr) \bigl(\overline{f}\gamma_\mu f\bigr)
+ \varepsilon_{\alpha\beta}^{fA}\,
\bigl(\overline{\nu}_\alpha\gamma^\mu P_L \nu_\beta\bigr) \bigl(\overline{f}\gamma_\mu \gamma^5 f\bigr)
\right\} ,
\label{eq:NSI}
\end{equation}
where $\alpha, \beta = e,\mu,\tau$ denote the neutrino flavors, $f = u,d$ label the up and down
quarks (we omit the leptonic NSIs, $f = e$, as they do not contribute to CE$\nu$NS),
summation over $\alpha, \beta$ and $f$ is implicit,
and $\varepsilon_{\alpha\beta}^{fV}$, $\varepsilon_{\alpha\beta}^{fA}$ are the vector
and axial-vector NSI couplings, respectively.
In this paper, we do not consider the axial-vector NSIs, because their contribution
to the CE$\nu$NS cross section depends on the nuclear spin and is therefore negligible
for heavy nuclei~\cite{AristizabalSierra:2018eqm}, similarly to the contribution of the
neutral current axial couplings in the SM~\cite{Barranco:2005yy,Miranda:2019skf}.

Non-standard neutrino-quark interactions can have a more general Lorentz structure than the one
of Eq.~(\ref{eq:NSI}). Namely, one may also consider scalar, pseudo-scalar and tensor couplings\footnote{The
tensor operators $\bigl(\overline{\nu}_\alpha \sigma^{\mu \nu} P_L \nu_\beta \bigr)
\bigl(\overline{q} \sigma_{\mu \nu} P_R q\bigr)$ vanish by Lorentz symmetry~\cite{Jenkins:2017jig},
hence do not appear in Eq.~(\ref{eq:GNI}).},
usually dubbed ``generalized neutrino interactions'' (GNIs) in the literature~\cite{Lindner:2016wff,AristizabalSierra:2018eqm,Bischer:2018zcz}
\begin{eqnarray}
\mathcal{L}_{\mathrm{GNI}}\, =\, -\, \sqrt{2}\,G_F\,
&& \left\{ \varepsilon_{\alpha\beta}^{qS}\,
\bigl(\overline{\nu}_\alpha P_L \nu_\beta \bigr)
\bigl(\overline{q} q\bigr)
+ \varepsilon_{\alpha\beta}^{qP}\,
\bigl(\overline{\nu}_\alpha P_L \nu_\beta \bigr)
\bigl(\overline{q} \gamma^5 q\bigr) \right.  \nonumber \\
&&  \left. + \,\varepsilon_{\alpha\beta}^{qT}\,
\bigl(\overline{\nu}_\alpha \sigma^{\mu \nu} P_L \nu_\beta \bigr)
\bigl(\overline{q} \sigma_{\mu \nu} P_L q\bigr) \right\} +\, {\rm h.c.}\, ,
\label{eq:GNI}
\end{eqnarray}
where $\sigma^{\mu \nu} = \frac{i}{2} \left[ \gamma^\mu, \gamma^\nu \right]$ and $q = u, d$.
Generally, NSIs are also included in the GNI Lagrangian. In the following, we will refer to
$\varepsilon_{\alpha\beta}^{qS}$, $\varepsilon_{\alpha\beta}^{qV}$ and $\varepsilon_{\alpha\beta}^{qT}$
as scalar, vector and tensor couplings, respectively
(we will not consider pseudo-scalar interactions, which give a negligible contribution to
CE$\nu$NS~\cite{Lindner:2016wff,AristizabalSierra:2018eqm}).
Note that hermiticity of the Lagrangian implies
$\varepsilon_{\beta\alpha}^{qV} = \left(\varepsilon_{\alpha\beta}^{qV}\right)^*$, while the matrices
of scalar and tensor couplings do not have any particular symmetry property if neutrinos are Dirac fermions.
If they are Majorana fermions, instead, one has $\varepsilon_{\beta\alpha}^{qS} = \varepsilon_{\alpha\beta}^{qS}$
and $\varepsilon_{\beta\alpha}^{qT} = - \varepsilon_{\alpha\beta}^{qT}$
(in particular, flavor-diagonal tensor couplings vanish for Majorana neutrinos).

In the presence of scalar, vector and tensor non-standard neutrino interactions,
the differential CE$\nu$NS cross section~(\ref{eq:CEvNS_SM})
for an incident neutrino of flavor $\alpha$ is modified to~\cite{Lindner:2016wff}
\begin{eqnarray}
  \frac{\mathrm{d}\sigma}{\mathrm{d}T} (E_\nu, T)\, =\ \frac{G_F^2 M}{\pi}\ F^2(|\vec{q}|^2)\!
  \sum_{\beta\, =\, e,\, \mu,\, \tau} && \left[\, \left| C^S_{\beta \alpha} \right|^2 \frac{MT}{8E_\nu^2}
  + \left| C^V_{\beta \alpha} + Q_{W, \alpha}\, \delta_{\alpha \beta}) \right|^2 \left ( 1-\frac{MT}{2E_\nu^2} \right ) \right.  \nonumber  \\
  & & \left. +\, 2 \left| C^T_{\beta \alpha} \right|^2 \left ( 1-\frac{MT}{4E_\nu^2} \right )  \right]\, ,
\label{eq:CEvNS_GNI}
\end{eqnarray}
where subleading terms of order $T/E_\nu \ll M T / E^2_\nu$ have been omitted\footnote{In particular,
we omit a scalar-tensor interference term in Eq.~(\ref{eq:CEvNS_GNI}), which is proportional
to $T/E_\nu$ and whose sign is different for neutrinos and antineutrinos~\cite{Lindner:2016wff}.}.
For an incident antineutrino of flavor $\alpha$, one should replace $C^S_{\beta \alpha} \to C^S_{\alpha \beta}$,
$C^V_{\beta \alpha} \to C^V_{\alpha \beta}$ and $C^T_{\beta \alpha} \to C^T_{\alpha \beta}$
in Eq.~(\ref{eq:CEvNS_GNI}).
The SM contribution is contained in the weak nuclear charge $Q_{W, \alpha}$, given by Eq.~(\ref{eq:weak_charge}),
while the  non-standar contributions are encapsulated in the coefficients $C^S_{\alpha \beta}$, $C^V_{\alpha \beta}$
and $C^T_{\alpha \beta}$, which are given by~\cite{AristizabalSierra:2018eqm}
\begin{equation}
C^S_{\alpha \beta}\, = \sum_{q = u,d} \left( Z\, \frac{m_p}{m_q}f_q^p + N\, \frac{m_n}{m_q}f_q^n \right)
  \varepsilon^{qS}_{\alpha \beta}\ ,
\label{eq:scalar}
\end{equation}
\vskip -.6cm
\begin{equation}
C^V_{\alpha \beta}\, =\, (2 Z + N)\, \varepsilon^{uV}_{\alpha \beta} + (Z + 2 N)\, \varepsilon^{dV}_{\alpha \beta}\ ,
\label{eq:vector}
\end{equation}
\vskip -.8cm
\begin{equation}
C^T_{\alpha \beta}\, = \sum_{q = u,d} \left( Z \delta_q^p + N \delta_q^n \right) \varepsilon^{qT}_{\alpha \beta}\ ,
\label{eq:tensor}
\end{equation}
For the numerical coefficients appearing in Eqs. (\ref{eq:scalar}) and (\ref{eq:tensor}), we adopt
the values given in Refs.~\cite{Hoferichter:2015dsa} and~\cite{Bhattacharya:2016zcn}, respectively:
\begin{equation}
f_u^p = 0.0208\, ,  ~~~~~ f_d^p =  0.0411\, ,  ~~~~~ f_u^n = 0.0189\, ,  ~~~~~ f_d^n =  0.0451\, ,
\end{equation}
\vskip -.8cm
\begin{equation}
\delta_u^p = 0.792\, ,  ~~~~~ \delta_d^p = -0.194\, , ~~~~~ \delta_u^n = -0.194\, , ~~~~~ \delta_d^n = ~0.792\, .
\end{equation}
As for quark masses, we use the central values given by the Particle Data Group~\cite{ParticleDataGroup:2022pth},
$m_u = 2.2\, \mbox{MeV}$ and $m_d = 4.7\, \mbox{MeV}$.
Given the large uncertainties associated with the light quark masses and with $f^N_q$ and $\delta^N_q$,
many authors provide constraints on the coefficients $C^S_{\alpha \beta}$ and $C^T_{\alpha \beta}$ rather than
on the Lagragian parameters $\varepsilon^{qS}_{\alpha \beta}$ and $\varepsilon^{qT}_{\alpha \beta}$.
In this paper, we choose instead to constrain the Lagragian parameters for fixed values of $m_q$, $f^N_q$
and $\delta^N_q$. It is straightforward to translate the current bounds and future sensitivity estimates
on $\varepsilon^{qS}_{\alpha \beta}$ and $\varepsilon^{qT}_{\alpha \beta}$ presented in the next sections
into constraints on $C^S_{\alpha \beta}$ and $C^T_{\alpha \beta}$.

Other authors, e.g. in Ref.~\cite{Lindner:2016wff,AristizabalSierra:2018eqm,Flores:2021kzl,Majumdar:2022nby,DeRomeri:2022twg},
use the symbols $C^q_S$, $C^q_V$, $C^q_T$ instead of $\varepsilon^{qS}$, $\varepsilon^{qV}$, $\varepsilon^{qT}$
in Eqs.~(\ref{eq:scalar}) to~(\ref{eq:tensor}), with depending on the normalization convention
either $C_V^q = \varepsilon^{qV}$~\cite{Majumdar:2022nby}
(with the CE$\nu$NS cross section given by Eq.~(\ref{eq:CEvNS_GNI}))
or $C_V^q = 2\varepsilon^{qV}$~\cite{Lindner:2016wff,AristizabalSierra:2018eqm,Flores:2021kzl,DeRomeri:2022twg}
(in which case one should replace $C^V_{\alpha \beta}$ by $\frac{1}{2}\, C^V_{\alpha \beta}$ in Eq.~(\ref{eq:CEvNS_GNI})).

\section{Experimental setup and detectors}
\label{sec:experiment}

In this section, we describe the SNS neutrino beam and the main characteristics of the three detectors
considered in this paper: the current CsI detector used by the COHERENT collaboration,
a future NaI detector similar to the one under deployment at the SNS~\cite{Hedges:2021mra}, and a cryogenic CsI detector proposed by the same collaboration~\cite{COHERENT:2023sol}.

\subsection{Neutrino beam and predicted number of CE$\nu$NS events}
\label{subsec:nu_beam}

At the SNS, the neutrinos used for CE$\nu$NS measurements arise from the decays at rest of the $\pi^+$
produced, together with neutrons and $\pi^-$'s, from the collision of high-energy protons on a mercury target
(in which the $\pi^-$ are absorbed before decaying).
The SNS neutrino beam therefore consists of three components: a prompt, mono-energetic $\nu_{\mu}$ component produced in the two-body decays of stopped positively charged pions ($\pi^+ \rightarrow \mu^+ + \nu_{\mu}$), and two delayed  $\bar{\nu}_{\mu}$ and $\nu_e$ components arising from the subsequent decays at rest of the antimuons ($\mu^+ \rightarrow e^+ + \bar{\nu}_{\mu} + \nu_e$).
These contributions can be analytically computed and are given by
\begin{equation}
\frac{dN_{\nu_\mu}}{dE_\nu}\, =\, \xi\, \delta\! \left(E_\nu - \frac{m_{\pi}^2-m_{\mu}^2}{2 m_{\pi}} \right)\, ,
\label{eq:flux:1}
\end{equation}
\vskip -.8cm
\begin{equation}
\frac{dN_{\bar{\nu}_\mu}}{dE_\nu}\, =\, \xi\, \frac{64 E_\nu^2}{m_{\mu}^3} \left(\frac{3}{4} - \frac{E_\nu}{m_{\mu}} \right)\, , \\
\label{eq:flux:2}
\end{equation}
\vskip -.8cm
\begin{equation}
\frac{dN_{\nu_e}}{dE_\nu}\, =\, \xi\, \frac{192 E_\nu^2}{m_{\mu}^3}\left(\frac{1}{2} - \frac{E_\nu}{m_{\mu}} \right)\, ,
\label{eq:flux:3}
\end{equation}
where $E_\nu$ is the neutrino energy and $\xi = r N_{\rm POT}/4\pi L^2$ is a normalization factor that depends on several experimental features: $N_{\rm POT}$, the number of protons on target (POT) throughout the operation time; $r$, the number of neutrinos per flavor produced by each POT; and $L$, the distance between the source and the detector.

Given the neutrino flux, the predicted number of CE$\nu$NS events in the nuclear recoil energy bin $[T_i, T_{i+1}]$
is given by, for a target material consisting of a single nucleus with mass $M$,
\begin{equation}
N^{th}_{i}\, =\, \mathcal{N}\int_{T_i}^{T_{i+1}}\! A(T)\, dT  \int_{0}^{T'_{\rm max}} \mathcal{R}(T,T')\, dT' \!
  \sum\limits_{\nu = \nu_e, \nu_{\mu}, \overline{\nu}_{\mu}} \int_{E_{\rm min}(T')}^{E_{\rm max}} dE_{\nu}\
  \frac{\mathrm{d} N_{\nu}}{\mathrm{d}\! E_{\nu}}(E_{\nu})\, \frac{\mathrm{d} \sigma }{\mathrm{d} T'}(E_{\nu},T')\ ,
\label{eq:CEvNS_events_NR_bins}
\end{equation}
where ${\mathrm{d} \sigma} / {\mathrm{d} T}$ is the relevant CE$\nu$NS cross section,
given by Eq.~(\ref{eq:CEvNS_SM}) or~(\ref{eq:CEvNS_GNI}),
and $\mathcal{N}=N_A M_{\textrm{det}}/M_{\textrm{mol}}$ is the number of nuclei in the detector,
which is computed from the detector mass $M_{\rm det}$
and the molar mass of the nucleus $M_{\rm mol}$, with $N_A$ the Avogadro constant.
The lower limit of the integral over $E_\nu$ is the minimal neutrino energy that can induce
a nuclear recoil of energy $T'$, $E_{\rm min}(T') = \sqrt{MT'/2}\,$, while the upper limit $E_{\rm max}$
is the maximum neutrino energy, which for the SNS beam is $52.8\, \mbox{MeV}$.
Notice that Eq. \eqref{eq:CEvNS_events_NR_bins} depends on both the real ($T'$) and the reconstructed ($T$)
nuclear recoil energies, with $\mathcal{R}(T,T')$ the smearing function, which may be different for each detector, and the upper limit of the integral over $T'$ is the maximal nuclear recoil energy that can be induced by a neutrino from the beam, $T'_{\rm max} = 2E_{\rm max}^2/M$. Finally, $A(T)$ is an acceptance function, which also depends on the detector.
If the target material consists of two nuclei, as is the case for the CsI and NaI detectors,
the total number of CE$\nu$NS events is the sum of the numbers of neutrino scatterings on each nucleus,
computed separately from Eq.~(\ref{eq:CEvNS_events_NR_bins}) with the appropriate values
of $M$ and $M_{\rm mol}$.

When the timing information of the experiment is available, one can also
bin the data in time intervals. In this case, the predicted number of events in the nuclear
recoil energy bin $i$ and time bin $j$ is given by
\begin{align}
  N_{ij}^{th}\ =\! \sum\limits_{\nu = \nu_e, \nu_{\mu}, \overline{\nu}_{\mu}}\!\! N_{\nu, i}^{th}\,
  \int_{t_j}^{t_{j+1}} f_\nu(t)\, \varepsilon_{t}(t)\, dt\, ,
\label{eq:CEvNS_events_NR_time_bins}
\end{align}
where $N_{\nu, i}^{th}$ is the predicted number of CE$\nu$NS events induced by the component $\nu$
($\nu = \nu_e, \nu_{\mu}, \overline{\nu}_{\mu}$) of the neutrino flux in the nuclear recoil energy bin $[T_i, T_{i+1}]$,
$f_\nu(t)$ is the time distribution of the neutrino flux component $\nu$ taken from Ref.~\cite{Picciau:2022xzi},
and $\varepsilon(t)$ is the timing efficiency provided in Ref.~\cite{Akimov:2021dab}.
To compute $N_{\nu, i}^{th}\,$, we use Eq.~(\ref{eq:CEvNS_events_NR_bins}) without performing
the sum over the contributions of the three components of the neutrino flux.

In our analysis, we use the data of the current CsI detector of the COHERENT collaboration,
as well as the expected future data from a NaI and cryogenic CsI detectors at the SNS.
The main characteristics of these detectors are described in the following subsections,
and the relevant properties of their target nuclei are given in Table~\ref{detectors_table1}.

\begin{table}[t]
\centering
\begin{tabular}{|c|c|c|c|c|}
\hline
\textbf{Target nucleus} & \textbf{ Z } & \textbf{ N } & \textbf{ Z/N } & \textbf{M (a.m.u)} \\ 
\hline
Cs              & 55       & 78         & 0.71         & 132.91   \\ 
\hline
I               & 53       & 74         & 0.72         & 126.90   \\ 
\hline
Na              & 11       & 12         & 0.92         &  22.99    \\ 
\hline
\end{tabular}
\caption{Main properties of the target nuclei used in the detectors considered in this work:
number of protons (Z) and neutrons (N),  proton-to-neutron ratio (Z/N) and mass in atomic units (M).}
\label{detectors_table1}
\end{table}

\subsection{Current CsI detector}
\label{subsec:CsI_detector}

The COHERENT collaboration performed CE$\nu$NS measurements with a 14.6 kg CsI detector located at 19.3~m from the neutrino source~\cite{Akimov:2021dab}. The accumulated data corresponds to $N_{POT} = 3.198\times10^{23}$, with a number of neutrinos per flavor of $r = 0.0848$~\cite{Akimov:2021dab}. These numbers are used to compute the neutrino fluxes though Eqs.~\eqref{eq:flux:1}, \eqref{eq:flux:2} and~\eqref{eq:flux:3}.
Following the COHERENT collaboration, we take the smearing function $\mathcal{R}(T,T')$
in Eq.~(\ref{eq:CEvNS_events_NR_bins}) to be the Gamma function given in Ref.~\cite{Akimov:2021dab}
(we refer the reader to Ref.~\cite{DeRomeri:2022twg} for details about the smearing procedure).
Finally, we take into account the timing information provided by the COHERENT collaboration
in our analysis and use Eq.~(\ref{eq:CEvNS_events_NR_time_bins}) to compute the predicted
number of events in each recoil energy and time bin.
In this way, we obtain an event spectrum in energy and time that reproduces the one presented in the appendix of Ref.~\cite{DeRomeri:2022twg}.

\subsection{Future NaI detector}
\label{subsec:NaI_detector}

The complete program of the COHERENT collaboration includes a NaI-based detector comprising several modules, each consisting of individual 7.7 kg NaI crystals \cite{Akimov:2022oyb}. In its final design, this detector is expected to have up to seven modules, each containing 63 crystals, giving a total mass $M_{\rm det} = 3395.7\, \mbox{kg}$, which we shall consider in this work. The detector is expected to be located at a distance $L = 22\, \mbox{m}$ from the neutrino source~\cite{Akimov:2022oyb}.  
By the time this detector is scheduled to be fully deployed, the SNS is expected to operate at an average power of 2 MW with a proton energy $E_p = 1.3\, \mbox{GeV}$, for which a value of $r = 0.13$ is predicted in Ref.~\cite{COHERENT:2021yvp}. Under these assumptions, we get a total number of protons on target (POT) of 1.73$\times10^{23}$ per calendar year of data taking,  assuming $\simeq 5000$ hours (208.3 days) of operation per year.  In this work, we consider three years of data taking with the SNS operating at a beam power of 2~MW with $E_p = 1.3\, \mbox{GeV}$, giving a total number of protons on target $N_{\rm POT} = 5.2 \times10^{23}$.

\begin{figure}[t]
\centering
\includegraphics[width=0.48\textwidth]{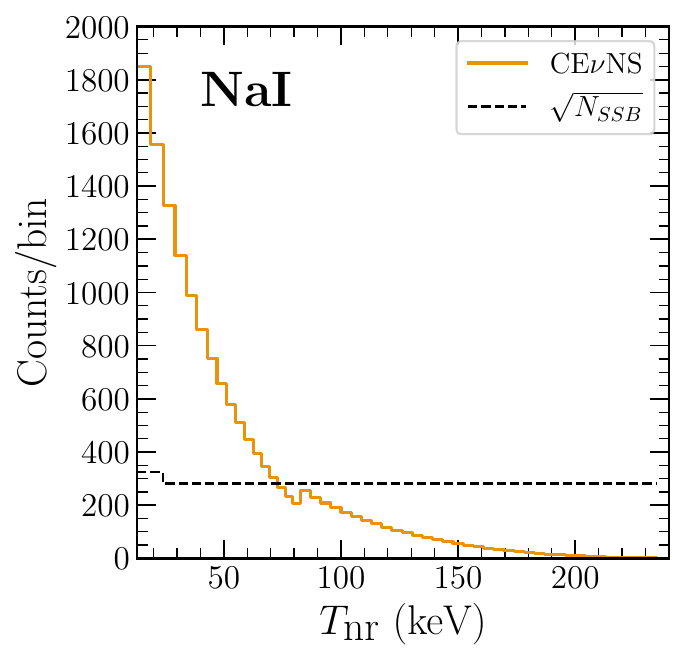}
\includegraphics[width=0.49\textwidth]{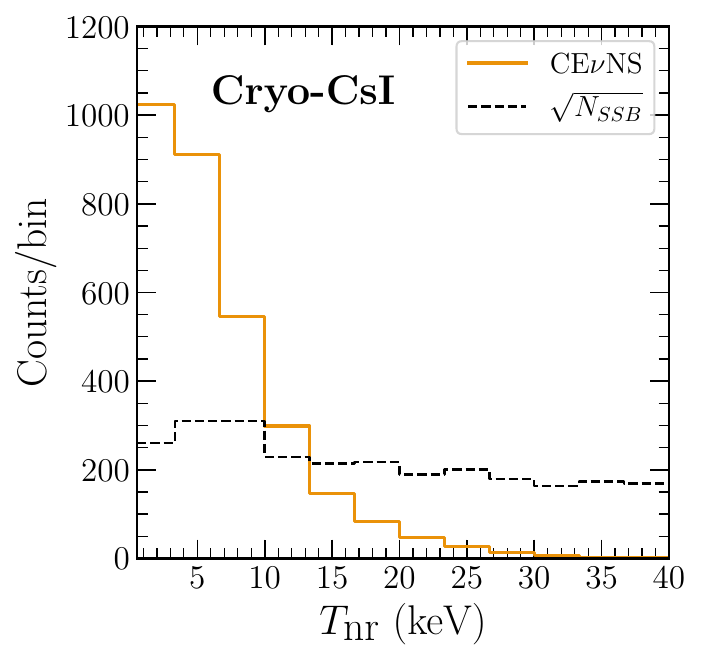}
\caption{Projected numbers of signal and background events in the SM as a function of the reconstructed nuclear recoil energy for the future NaI (left) and Cryo-CsI (right) detectors considered in this paper, assuming 3 years of data taking.
The signal spectra include the contributions of the three components of the SNS neutrino beam ($\nu_e$, $\nu_{\mu}$ and $\overline{\nu}_{\mu}$).
The expected steady-state background is represented by the black dashed curve, which shows the square root of the actual number of background events for NaI.}
\label{Events-NaI}
\end{figure}

Even though the NaI molecule contains two nuclei, the detector will be only sensitive to Na recoils.
Indeed, due to a low quenching factor ${\rm QF (I)} \leq 8 \%$, the nuclear recoil energy threshold for iodine is
such that only a tiny fraction of the iodine-induced CE$\nu$NS events can be detected\footnote{As can be seen from Fig.~26 of Ref.~\cite{Cintas:2024pdu}, the quenching factor for iodine is smaller than $8 \%$ over the recoil energy range $[0, 80]\ {\rm keV}_{\rm nr}$. The NaI detector has an electron-equivalent energy threshold of $3\, {\rm keV}_{\rm ee}$, which, for ${\rm QF (I)} = 8 \%$, corresponds to a nuclear recoil energy threshold of $37.5\, {\rm keV}_{\rm nr}$, not far below the maximal iodine nucleus recoil energy that a neutrino from the SNS beam can induce ($T'_{\rm max} {\rm (I)} \approx 47\, {\rm keV}_{\rm nr}$).}.
We can therefore safely neglect them.
To compute the expected number of (Na-induced) CE$\nu$NS events in the future NaI detector,
we assume a conservative constant acceptance of 80\%
and a Gaussian smearing function $\mathcal{R}(T,T')$
with an energy-dependent resolution $\sigma(T) = \eta\sqrt{TT_{th}}$, where\footnote{This choice
reproduces the energy resolution given in~\cite{Hedges_M7_2018} to a good approximation.}
$\eta = 0.14$ and $T_{th} = 13\, \mbox{keV}_{\rm nr}$ is the sodium nucleus recoil energy threshold
of the detector~\cite{Hedges:2021mra}.
We thus obtain the event spectrum shown in the left panel of Fig.~\ref{Events-NaI},
where the width of the recoil energy bins corresponds to $1\, \mbox{keV}_{\rm ee}$~\cite{Hedges:2021mra}.
In order to convert an electron recoil energy of $1\, \mbox{keV}_{\rm ee}$ into the equivalent nuclear recoil energy (i.e., the nuclear recoil energy giving the same number of photoelectrons as an electron recoil energy of $1\, \mbox{keV}$),  we rely on
the recent Na quenching factor measurements of Ref.~\cite{Cintas:2024pdu}.
Namely, for nuclear recoil energies between 13 keV$_{\textrm{nr}}$ and 82 keV$_{\textrm{nr}}$,
we fit the QF data from Fig.~22 of Ref.~\cite{Cintas:2024pdu}
as a linear function of the nuclear recoil energy.
For energies above 82 keV$_{\textrm{nr}}$ and up to $T_{\rm max} ({\rm Na}) \approx 259\, \mbox{keV}_{\rm nr}$,
for which no measurements are available,
we use a constant quenching factor of 0.23 (estimated from Fig.~4.12 of Ref.~\cite{Hedges:2021mra}), resulting in a constant bin width of $4.33\, \mbox{keV}_{\rm nr}$. This is larger than the width of the previous bins, thus explaining the little bump at 82 keV$_{\textrm{nr}}$ in the histogram on the left panel of Fig.~\ref{Events-NaI}.
On the same plot, we show the expected steady-state background, which we assumed to be flat
and equal to its average value of $300$ ckkd (counts per keV$_{ee}$ per kg per day)~\cite{Hedges:2021mra},
except in the first two bins where the background is larger. There, we took a conservative value of 400 ckkd.
In practice, the number of background events is reduced by a factor of 8000~\cite{Hedges:2021mra}
taking into account the fact that the SNS neutrino beam is pulsed. This reduction factor has been applied
in the left panel of Fig.~\ref{Events-NaI}.

\subsection{Future cryogenic CsI detector}
\label{subsec:Cryo-CsI_detector}

In addition to the NaI detector, the COHERENT program includes a future cryogenic CsI (Cryo-CsI)
detector, with expected improved threshold and detection efficiency with respect to the CsI detector
used for the first CEvNS observation. In this paper, we aim to explore the sensitivity of this future Cryo-CsI detector to GNIs and to show how the combination with the future NaI detector can help to break certain degeneracies in the parameter space. 
For the experimental design, we follow Ref.~\cite{COHERENT:2023sol} and assume a 10 kg cryogenic  CsI detector located at 19.3~m from the neutrino source, like the current CsI detector.
Regarding SNS operations, we make the same assumptions as for the NaI detector and consider a total number of protons on target of 1.73$\times10^{23}$ per year of data taking. In Ref.~\cite{COHERENT:2023sol}, it is mentioned that a detector threshold of 4 photoelectrons (PE) is expected, which we also assume here. To convert this threshold into a nuclear recoil energy, we consider a constant quenching factor of 15$\%$ and a light yield of 50 PE/keV$_{\textrm{ee}}$, as indicated in the same reference. Under these assumptions,  a 4 PE threshold corresponds to a nuclear recoil energy threshold $T_{th} = 0.533\, {\rm keV}_{\rm nr}$. 
As for the detector efficiency, we consider the same efficiency function as the one provided by the COHERENT collaboration for the current CsI detector (see the supplementary material of Ref.~\cite{Akimov:2021dab}), but with a shift of 8 PE starting at 4 PE in order to account for the fact that a cryogenic detector will improve the efficiency at lower threshold energies.
Finally, we neglect smearing between the true recoil energy and the observed number of photoelectrons, as it is estimated to be minimal in Ref.~\cite{COHERENT:2023sol}, due to the high expected light yield.
With these assumptions, we are able to reproduce the expected CE$\nu$NS event rate shown in Fig.~3 of Ref.~\cite{COHERENT:2023sol}.
In the right panel of Fig.~\ref{Events-NaI}, we show the expected number of signal events and
steady-state background, using the background model of Ref.~\cite{COHERENT:2023sol} and applying
a reduction factor of 2000, as indicated in Ref.~\cite{Ding:2024mra}. Note that we adopted the same bin width
of 25 PE used in Fig.~2 of Ref.~\cite{COHERENT:2023sol} to display their steady-state background model.

\section{Analysis procedure}
\label{sec:analysis}

For the analysis of the current CsI detector data, we follow the same procedure as Ref.~\cite{DeRomeri:2022twg}
and use the Poissonian  $\chi^2$ function
\begin{equation}
  \chi^2 (\kappa)\, =\, \underset{\xi}{\rm min} \left\{ 2\sum_{i, j} \left [ N_{ij}(\kappa, \xi) - \tilde{N}_{ij} + \tilde{N}_{ij}\ln\left ( \frac{\tilde{N}_{ij}}{N_{ij}(\kappa,\xi)} \right )\right ] + \sum_{m=1}^2 \frac{\xi^2_{sig, m}}{\sigma_{\xi_{sig,m}}^2} + \sum_{k=1}^3\frac{\xi^2_{bg, k}}{\sigma_{\xi_{bg,k}}^2} \right\} ,
\label{eq:chi:CsI}
\end{equation}
where $i, j$ run over the nuclear recoil energy and time bins, respectively, $\tilde{N}_{ij}$ is the experimental number of events
in bins $(i,j)$, $N_{ij}(\kappa,\xi)$ the predicted number of events in the same bins in the presence of GNIs,
$\kappa$ stands for the set of GNI parameters under test, and $\xi$
is the set of nuisance parameters over which we minimize the expression within braces.
$N_{ij}(\kappa,\xi)$ is computed as
\begin{eqnarray}
N_{ij}(\kappa,\xi) =&& ~(1+\xi_{sig,1})N_{ij}^{sig}(\kappa,\xi_{sig,2}, \xi_1,\xi_2) + (1+\xi_{bg,1})N_{ij}^{BRN}(\xi_1) \nonumber  \\
  & & +\, (1+\xi_{bg,2})N_{ij}^{NIN}(\xi_1)+ (1+\xi_{bg,3})N_{ij}^{SSB},
\end{eqnarray}
where $N_{ij}^{sig}$ is the number of signal events, and $N_{ij}^{BRN}$, $N_{ij}^{NIN}$ and $N_{ij}^{SSB}$
are the numbers of beam-related neutron (BRN), neutrino-induced neutron (NIN) and steady-state (SSB)
background events, respectively. The signal nuisance parameter $\xi_{sig,1}$ accounts for the detector efficiency,
neutrino flux, and quenching factor (QF) normalizations, while $\xi_{sig,2}$ is related to the nuclear radius.
The corresponding uncertainties are $\sigma_{\xi_{sig,1}} = 11.45\%$ and $\sigma_{\xi_{sig,2}} = 5\%$~\cite{DeRomeri:2022twg}.
As for the nuisance parameters $\xi_{bg,k}$, they are associated with the different sources of background:
BRN ($\xi_{bg,1}$), NIN ($\xi_{bg,2}$) and SSB ($\xi_{bg,3}$), with corresponding uncertainties
$\sigma_{\xi_{bg,1}} = 25\%$, $\sigma_{\xi_{bg,2}} = 35\%$ and $\sigma_{\xi_{bg,3}} = 2.1\%$~\cite{Akimov:2021dab}.
Following Ref.~\cite{DeRomeri:2022twg}, we include in Eq.~\eqref{eq:chi:CsI} two additional
nuisance parameters $\xi_1$ and $\xi_2$, with no penalization term, which account for deviations
in the uncertainty on the CE$\nu$NS detection efficiency and on beam timing, respectively.
By minimizing over all nuisance parameters, we obtain allowed regions at a given confidence level
for the GNI parameters under test.

To assess the sensitivity of the future NaI and Cryo-CsI detectors to GNI parameters,
we follow Ref.~\cite{Chatterjee:2022mmu} and consider the $\chi^2$ function
\begin{equation}
\chi^2(\kappa)\, =\, \underset{\xi}{\rm min}\left[\sum\limits_{i}2\left\{N_i(\kappa, \xi) - \tilde{N_{i}} + \tilde{N_{i}}\,{\rm ln}\left(\frac{\tilde{N_{i}}}{N_i(\kappa, \xi)}\right) \right\} + \left(\frac{\xi_{sig}}{\sigma_{sig}}\right)^2 + \left(\frac{\xi_{bg}}{\sigma_{bg}}\right)^2 \right] ,
\label{eq:chi:NaI}
\end{equation}
where we have divided the data in recoil energy bins labelled by $i$.
Since we are dealing with simulated data, we compare the theoretical predictions in the absence
and in the presence of GNIs. Hence, we take
\begin{eqnarray}
\tilde{N_i} = N_i^{sig}(SM) + N_i^{bg}\, ,
\end{eqnarray}
where $N_i^{sig}(SM)$ represents the number of signal events in bin $i$ predicted by the SM
and $N_i^{bg}$ is the simulated steady-state background described in 
Subsections~\ref{subsec:NaI_detector} and~\ref{subsec:Cryo-CsI_detector}, while
\begin{equation}
N_i(\kappa, \xi) = N_i^{sig}(\kappa) (1+\xi_{sig}) + N_i^{bg}(1+\xi_{bg})\, ,
\end{equation}
where $N_i^{sig}(\kappa)$ is the predicted number of signal events in bin $i$ in the presence of GNIs,
and the nuisance parameters $\xi_{sig}$, $\xi_{bg}$ account for the signal and background normalizations,
respectively.
For the associated systematic uncertainties,
we conservatively assume $\sigma_{sig} = 10\%$ and $\sigma_{bg} = 5\%$ for both detectors.
This choice is motivated by the known values of the uncertainties for the current CsI detector.
Notice that, due to the absence of detailed information on the signal and background systematic uncertainties
for the NaI and Cryo-CsI detectors, we only  consider two nuisance parameters.

Finally, when we perform a combined analysis of future Cryo-CsI and NaI data,
we assume the signal and background systematic uncertainties to be fully uncorrelated.
In practice, this means that we add the two $\chi^2$ functions for each detector, with independent
nuisance parameters ($\xi^{\rm Cryo-CsI}_{sig}$, $\xi^{\rm Cryo-CsI}_{bg}$) and
($\xi^{\rm NaI}_{sig}$, $\xi^{\rm NaI}_{bg}$), respectively.
In general, correlations between systematic errors result in stronger constraints
on new physics parameters (see for instance Ref.~\cite{Galindo-Uribarri:2020huw}).
However, the level of correlation between uncertainties in future experiments
is difficult to estimate, so we take the conservative approach of assuming fully uncorrelated uncertainties
between the two detectors.

\section{Expected sensitivities to GNI parameters}
\label{sec:results}

In this section, we study the sensitivities to GNI parameters\footnote{From now on, we will
use the generic term ``GNI parameters'' to collectively refer to vector, scalar and tensor couplings,
while vector couplings will also be called ``NSI parameters''.} of future $\rm NaI$ and cryogenic CsI detectors
with the characteristics described in Subsection~\ref{subsec:NaI_detector},
and compare their combined expected sensitivities 
with the constraints set by the current COHERENT CsI detector, which we refer to as CsI (2021).
As discussed in Subsection~\ref{subsec:GNIs}, we consider only scalar, vector and tensor interactions
and denote the corresponding couplings by $\varepsilon^{qS}_{\alpha \beta}$, $\varepsilon^{qV}_{\alpha \beta}$
and $\varepsilon^{qT}_{\alpha \beta}$, respectively, with $\alpha, \beta = e, \mu, \tau$ and $q = u, d$.

Since the SNS neutrino beam consists only of electron neutrinos, muon neutrinos and muon antineutrinos,
not all of these parameters can be constrained by CE$\nu$NS measurements.
Vector GNIs, also known as NSIs, are subject  to the hermiticity condition
$\varepsilon_{\beta\alpha}^{qV} = \left(\varepsilon_{\alpha\beta}^{qV}\right)^*$.
As a result, there are 6 independent parameters for each quark flavor, but only 5 of them can be constrained at the SNS:
two real flavor-diagonal couplings, $\varepsilon^{qV}_{ee}$ and $\varepsilon^{qV}_{\mu\mu}$,
and three complex flavor off-diagonal couplings,
$\varepsilon^{qV}_{e\mu}$, $\varepsilon^{qV}_{e\tau}$ and $\varepsilon^{qV}_{\mu\tau}$.
In this paper, we study the sensitivities of the NaI and Cryo-CsI detectors
to $\varepsilon^{qV}_{ee}$,
$\varepsilon^{qV}_{\mu\mu}$ and $\varepsilon^{qV}_{e\mu}$ ($q=u,d$), leaving aside the other
off-diagonal couplings, which are less constrained\footnote{This
is due to the fact that $\varepsilon^{qV}_{e\mu}$ can induce coherent scatterings
for all three components of the SNS neutrino flux, while only $\nu_e$ (resp. $\nu_\mu$ and
$\nu_{\bar \mu}$) can scatter off a nucleus via $\varepsilon^{qV}_{e\tau}$ (resp. $\varepsilon^{qV}_{\mu\tau}$).}.

For scalar and tensor GNIs, the number of independent parameters depends on whether neutrinos
are Dirac or Majorana fermions. In this paper, we assume that they are Majorana fermions,
both for simplicity and because this possibility is better motivated from a theoretical point of view.
Scalar couplings are therefore symmetric, and for each quark flavor, 5 out of the 6 independent
ones can be constrained by CE$\nu$NS measurements at the SNS.
However, due to the absence of interference between the scalar and SM contributions
to the CE$\nu$NS cross section, the same constraints apply to $\varepsilon^{qS}_{ee}$ and
$\varepsilon^{qS}_{e\tau}$, as well as to $\varepsilon^{qS}_{\mu\mu}$ and $\varepsilon^{qS}_{\mu\tau}$.
It is therefore sufficient to study the sensitivities of the CsI and NaI detectors to the scalar couplings
$\varepsilon^{qS}_{ee}$, $\varepsilon^{qS}_{\mu\mu}$ and $\varepsilon^{qS}_{e \mu}$ ($q=u,d$).
Finally, tensor couplings are antisymmetric for Majorana neutrinos, leaving only 3 indepedent
parameters $\varepsilon^{qT}_{e\mu}$, $\varepsilon^{qT}_{e\tau}$ and $\varepsilon^{qT}_{\mu\tau}$
for each value of $q$, all of which can induce scatterings of SNS neutrinos off nuclei.

Throughout this paper, we assume all GNI couplings to be real.
The upper bounds on these parameters obtained from the currently available CsI data and the expected sensitivity
of the future NaI and Cryo-CsI detectors are determined following the numerical procedure described in Section~\ref{sec:analysis}.

\subsection{Expected sensitivities to individual GNI parameters}
\label{subsec:single_GNI}

\begin{figure}
\centering
\includegraphics[width=0.45\textwidth]{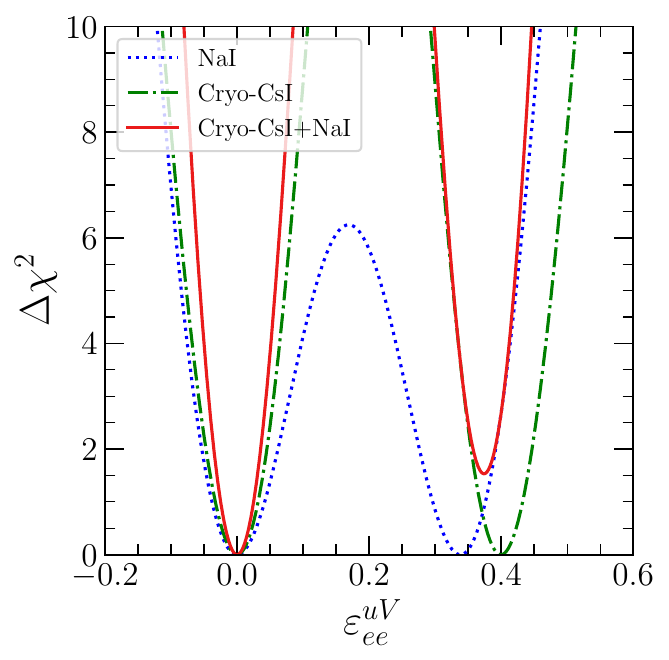}
\includegraphics[width=0.45\textwidth]{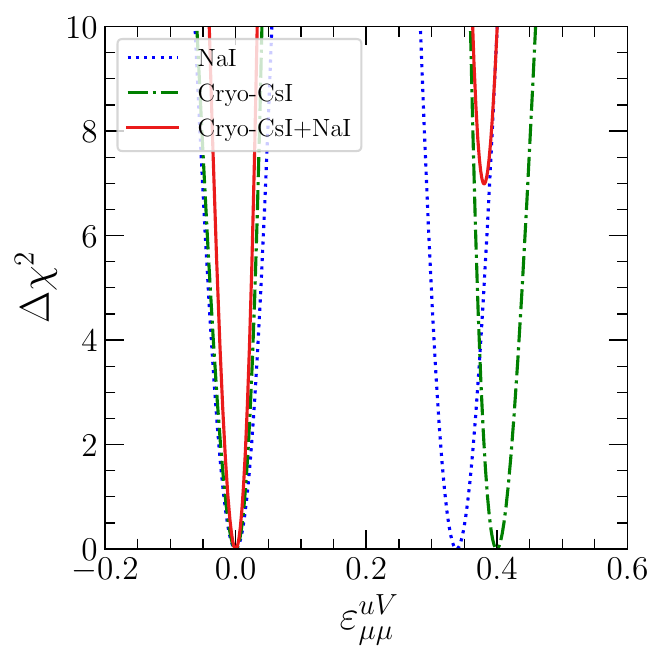}
\caption{Expected one-dimensional $\Delta \chi^2$ profiles 
for the vector couplings $\varepsilon^{uV}_{ee}$ (left panel) and $\varepsilon^{uV}_{\mu\mu}$ (right panel).
The blue dotted curves (resp. the green dash-dotted curves) represent the expected sensitivities
of the $\rm NaI$ detector described in Subsection~\ref{subsec:NaI_detector}
(resp. the Cryo-CsI detector described in Subsection~\ref{subsec:Cryo-CsI_detector}),
assuming 3 years of data taking at the SNS.
The solid red curves correspond to the combined ${\rm NaI} + {\rm Cryo}$-CsI analysis.}
\label{fig:CryoNaI1D}
\end{figure}

We first consider the expected sensitivities 
of the future NaI and Cryo-CsI detectors to individual parameters, i.e. we consider
a single nonvanishing coupling at a time. Note that in this case, the assumption of real couplings
does not mean any loss of generality, as the CE$\nu$NS cross section only depends on the moduli
of the GNI parameters (except for flavor-diagonal vector couplings, which are real and can interfere with the SM contribution).

Fig.~\ref{fig:CryoNaI1D} shows the expected sensitivities of the future detectors
to the flavor-diagonal vector couplings $\varepsilon^{uV}_{ee}$ (left panel) and $\varepsilon^{uV}_{\mu\mu}$ (right panel)
after 3 years of data taking at the SNS, assuming conservatively $\sigma_{sig} = 10\%$ and $ \sigma_{bg} = 5\%$.
In both panels, the blue dotted curves correspond to the NaI detector, and the green dash-dotted curves to the Cryo-CsI detector.
Two degenerate minima can be observed in the $\Delta\chi^2$ profiles, as a result of the interference between the SM
and NSI contributions to the CE$\nu$NS cross section.
Indeed, Eq.~(\ref{eq:CEvNS_GNI}) contains a term proportional to
(where we have set $\varepsilon^{dV}_{\alpha \alpha} = 0$ and neglected a term suppressed by $g^p_V \ll 1$)
\begin{equation}
  \left| C^V_{\alpha \alpha} + Q_{W, \alpha} \right|^2\,
    \simeq\, N^2 \left|\, \left( 2\, \frac{Z}{N} + 1 \right) \varepsilon^{uV}_{\alpha \alpha}
    - \frac{1}{2}\, \right|^2 ,
\label{eq:weak_charge_NSIs}
\end{equation}
implying the existence of a second $\Delta \chi^2$ minimum at $\varepsilon^{uV}_{\alpha \alpha} \simeq 1 / (2 Z/N + 1)$,
i.e. $\simeq 0.34$ for NaI and $\simeq 0.40$ for CsI (taking into account the fact
that the NaI detector is only sensitive to Na recoils, and that Cs and I have approximately the same $Z/N$).
The sharp rise of the $\Delta \chi^2$ functions between the two minima
is due to the fact that for $\varepsilon_{\alpha \alpha}^{uV} \simeq 1 / 2 (2 Z/N + 1)$,
the NSI and SM contributions to $\nu_\alpha$/$\bar \nu_\alpha$ scatterings completely cancel out,
resulting in a reduction of the predicted number of CE$\nu$NS events
by roughly 2/3 for $\alpha = \mu$ and 1/3 for $\alpha = e$, in strong tension with data.

Fig.~\ref{fig:CryoNaI1D} also shows the expected sensitivity to
$\varepsilon_{ee}^{uV}$ and $\varepsilon_{\mu\mu}^{uV}$ from a combined analysis
of the data of the future NaI and Cryo-CsI detectors (solid red curve),
assuming 3 years of data taking for each detector.
The advantage of such a combination is to lift the degeneracy between the two minima of the $\Delta \chi^2$ function:
since the location of the second minimum is different for NaI and Cryo-CsI
(as a consequence of the different proton-to-neutron ratios of their material targets),
it is difficult to fit SM-like experimental data in both detectors
with a nonzero value of $\varepsilon^{uV}_{\alpha \alpha}$.
In particular, the second $\Delta \chi^2$ minimum
is disfavoured at more than 99\%~C.L. for $\varepsilon_{\mu\mu}^{uV}$
(but only at a bit more than $1 \sigma$ for $\varepsilon_{ee}^{uV}$, due to the smaller number
of $\nu_e$-induced CE$\nu$NS events).

\begin{figure}
\centering
\includegraphics[width=0.32\textwidth]{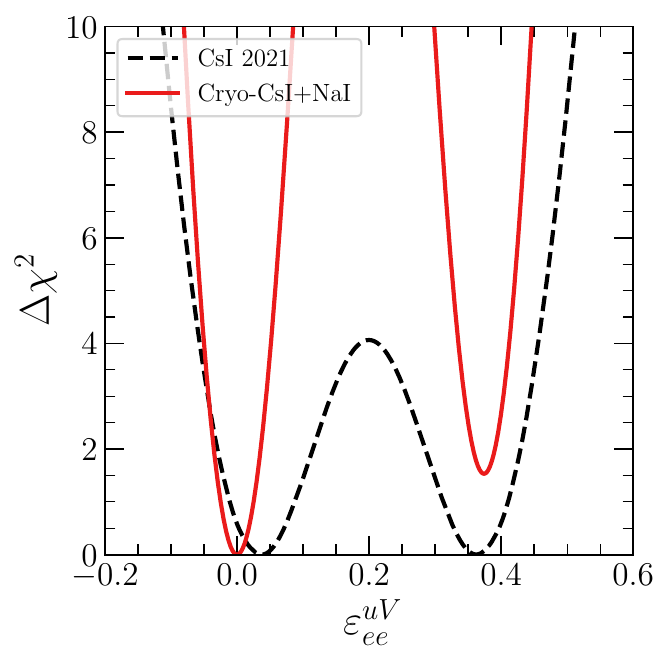}
\includegraphics[width=0.32\textwidth]{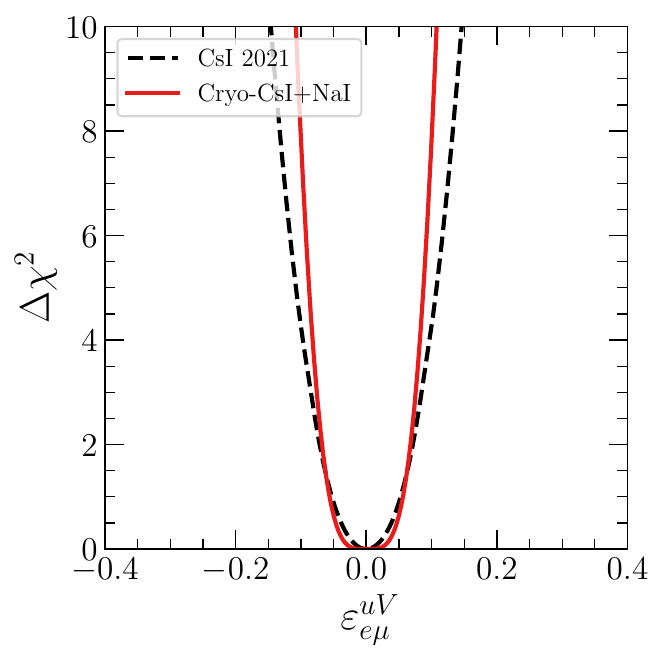}
\includegraphics[width=0.32\textwidth]{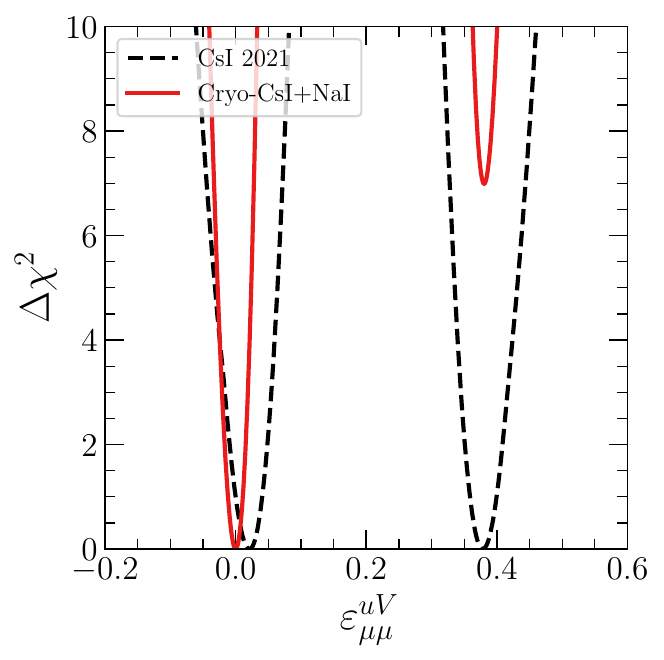}
\includegraphics[width=0.32\textwidth]{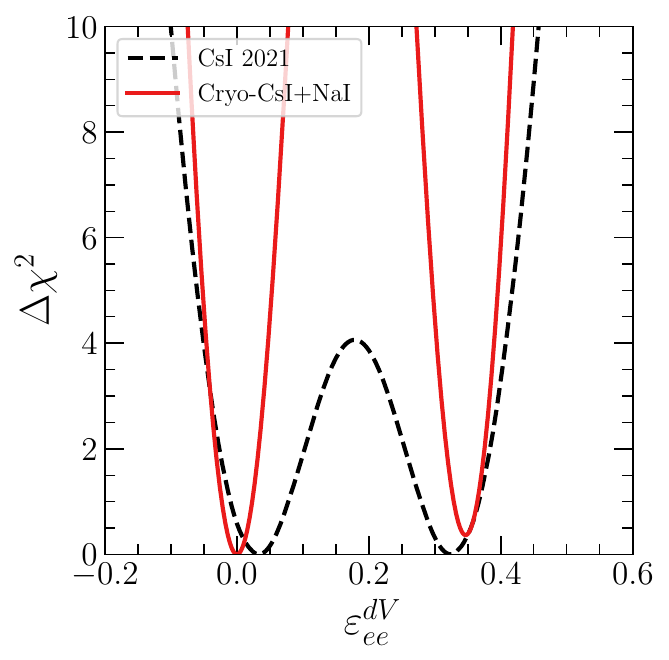}
\includegraphics[width=0.32\textwidth]{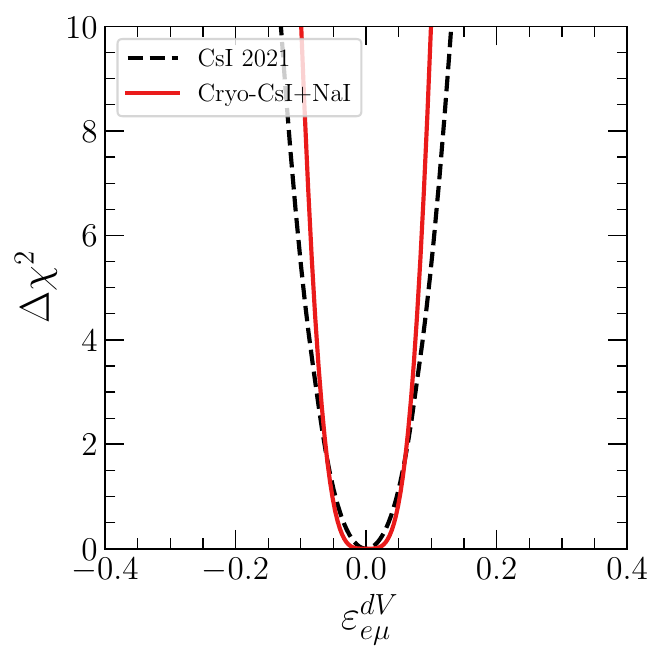}
\includegraphics[width=0.32\textwidth]{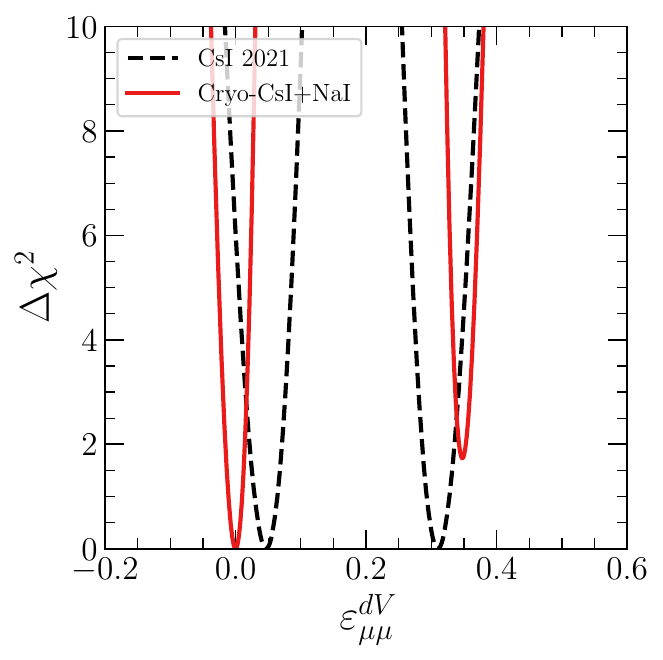}
\caption{One-dimensional $\Delta \chi^2$ profiles
for the vector coupling $\varepsilon^{qV}_{ee}$ (left panels), $\varepsilon^{qV}_{e\mu}$ (middle panels)
and $\varepsilon^{qV}_{\mu\mu}$ (right panels), with $q=u$ in the upper panels and $q=d$ in the lower panels.
The black dashed curves are obtained using the currently available data of the COHERENT CsI detector.
The solid red curves represent the expected sensitivities of the combination of the future NaI and Cryo-CsI detectors,
assuming 3 years of data taking at the SNS for each detector.
The flavor off-diagonal NSI parameter $\varepsilon^{qV}_{e\mu}$ is assumed to be real.}
\label{1D_proj_vector-new2}
\end{figure}

Having demonstrated the interest of performing a combined analysis of the data from
the future NaI and Cryo-CsI detectors, we study in the following the expected sensitivities to GNI
parameters of this detector combination, and we compare them with the constraints
obtained from the currently available CsI data.
The results for vector couplings are shown in Fig.~\ref{1D_proj_vector-new2},
where the $\Delta \chi^2$ functions defined in Section~\ref{sec:analysis} are plotted against the parameter
assumed to be nonvanishing (from left to right: $\varepsilon^{qV}_{ee}$, $\varepsilon^{qV}_{e\mu}$
and $\varepsilon^{qV}_{\mu\mu}$, with $q=u$ in the upper panels and $q=d$ in the lower panels).
The black dashed curves show the constraints obtained from the currently available (2021) COHERENT CsI data\footnote{Note that for the CsI detector, the first minimum of the $\Delta \chi^2$ function
for flavor-diagonal NSI parameters is not located at  $\varepsilon_{\alpha\alpha}^{qV} = 0$
($\alpha = e, \mu$; $q = u, d$). This is due to the fact that we are analysing real data,
which does not exactly match the SM prediction.} and are consistent
with the results obtained in Ref.~\cite{DeRomeri:2022twg}.
The solid red curves represent the expected
sensitivities of the combination of the future NaI and Cryo-CsI detectors after 3 years of data taking,
assuming again $\sigma_{sig} = 10\%$ and $ \sigma_{bg} = 5\%$.
Table~\ref{const_table_vector} summarizes the  $90\%$ C.L. and $2\sigma$ expected sensitivities
to vector couplings of the ${\rm NaI} + {\rm Cryo}$-CsI detector combination, as well as
the constraints on the same parameters from the COHERENT CsI 2021 data.

\begin{table}
\newcommand{\mc}[3]{\multicolumn{#1}{#2}{#3}}
\newcommand{\mr}[3]{\multirow{#1}{#2}{#3}}
\centering
\begin{adjustbox}{width=1\textwidth}
\begin{tabular}{|c|c|c|c|c|c|c|c|}
\hline
 \mr{2}{*}{Detector} & \mr{2}{*}{$\hskip .2cm q \hskip .2cm$} & \mc{2}{c|}{$\varepsilon^{qV}_{ee}$} &  \mc{2}{c|}{$\varepsilon^{qV}_{e\mu}$} &  \mc{2}{c|}{$\varepsilon^{qV}_{\mu\mu}$}  \\ 
 \cline{3-4}
 \cline{5-6}
 \cline{7-8}
  &  & $90\%\,\rm C.L.$  & $2\sigma\,\rm C.L.$ & $90\%\,\rm C.L.$ & $2\sigma\,\rm C.L.$  &$90\%\,\rm C.L.$ & $2\sigma\,\rm C.L.$\\
\hline
\hline
\mr{4}{*}{CsI (2021)}   &  \mr{2}{*}{$u$}  & $\left[-0.04, 0.134\right]\,$  & $\left[-0.056, 0.186\right]\,  $ &  $\left[-0.081, 0.081\right]\,$     & $\left[-0.097, 0.097\right]\,$ & $\left[-0.015, 0.053\right]\,$ & $\left[-0.024, 0.060\right]\,$     \\ 
  &		& $\cup\left[0.266, 0.440\right]\,$ &  $\cup\left[0.212, 0.456\right]\,$  &   $$ 	& $$   &$\cup\left[0.346, 0.414\right]\,$ & $\cup\left[0.34, 0.424\right]\,$\\
  & \mr{2}{*}{$d$}  & $\left[-0.035, 0.120\right]\, $ &$\left[-0.05, 0.168\right]\, $ &	  $\left[-0.073, 0.073\right]\,	$  &  $\left[-0.087, 0.087\right]\, 	$  &  $\left[-0.016, 0.075\right]\,	$ &  $\left[-0.010, 0.081\right]\,	$  \\
 &		& $\cup\left[0.237, 0.393\right]\, $  	&   $\cup\left[0.190, 0.408\right]\, $	&	$$	&	$$	 &  $\cup\left[0.282, 0.340\right]\,$ &  $\cup\left[0.276, 0.347\right]\,$    \\ 
\hline
\mr{4}{*}{${\rm NaI} + {\rm Cryo}$-CsI}   &  \mr{2}{*}{$u$}  & $\left[-0.041, 0.042\right]\,$  & $\left[-0.05, 0.051\right]\,  $ &  $\left[-0.073, 0.073\right]\,$     & $\left[-0.082, 0.082\right]\,$ &   $\left[-0.019, 0.018\right]\,$  &  $\left[-0.024, 0.021\right]\,	$    \\ 
 &		&  $\cup \left[0.347, 0.4\right]\,$   & $\cup\left[0.335, 0.412\right]$	  &   $$ 	& $$   &   & \\

  & \mr{2}{*}{$d$}  & $\left[-0.037, 0.038\right]\, $ &$\left[-0.046, 0.047\right]\, $ &	  $\left[-0.067, 0.067\right]\,	$  &  $\left[-0.075, 0.075\right]\,$  &  $\left[-0.018, 0.016\right]\,	$ &  $\left[-0.022, 0.020\right]\,$ \\
 &		& $\cup\left[0.311, 0.385\right]$  	&   $\cup\left[0.302, 0.389\right]$	&	$$	&		$$	&  $\cup\left[0.338, 0.358\right]\,$ & $\cup\left[0.333, 0.363\right]\,	$    \\
\hline
\end{tabular}
\end{adjustbox}
\caption{Sensitivities to the vector couplings
$\varepsilon^{qV}_{ee}$, $\varepsilon^{qV}_{e\mu}$ and $\varepsilon^{qV}_{\mu\mu}$ ($q=u,\,d$) at the $90\%$
and $2\sigma$ confidence levels with 1 degree of freedom (i.e. $\Delta\chi^2 \leq 2.71$ and $\Delta\chi^2 \leq 4$, respectively).
For the $\rm CsI$ detector, the intervals actually correspond to the ranges of parameter values
allowed by the currently available data,
while for the combination of the future NaI and Cryo-CsI detectors,
they represent the expected sensitivities assuming 3 years of data taking at the SNS
for each detector.
The flavor off-diagonal NSI parameter $\varepsilon^{qV}_{e\mu}$ is assumed to be real.}
\label{const_table_vector}
\end{table}

As can be seen from Fig.~\ref{1D_proj_vector-new2} and Table~\ref{const_table_vector},
a clear improvement of the constraints on flavor-diagonal vector couplings is expected from
the combination of the future NaI and Cryo-CsI detectors,
already with 3 years of data taking.
This is partly due to the larger statistics of these detectors.
However, the most noticeable feature, namely the fact that the second minimum of the $\Delta \chi^2$ function
tends to be disfavoured (most notably for the parameter $\varepsilon_{\mu\mu}^{uV}$,
and to a lesser extent for $\varepsilon_{ee}^{uV}$ and $\varepsilon_{\mu\mu}^{dV}$),
cannot be explained by statistics alone, but it is a consequence of the different proton-to-neutron ratios
of the Na and CsI targets,
as explained before. A similar result can be obtained by suitably choosing combinations of the detectors
proposed for CE$\nu$NS measurements at the European Spallation Source~\cite{Baxter:2019mcx},
such as ${\rm CsI} + {\rm Si}$ and ${\rm Xe} + {\rm Si}$, as shown in Ref.~\cite{Chatterjee:2022mmu}.

For flavor off-diagonal vector couplings (middle panels of Fig.~\ref{1D_proj_vector-new2}), only a small improvement on the current constraints
(mainly visible at $\gtrsim 3\sigma$ confidence level) is expected from the combination
of the future NaI and Cryo-CsI detectors, in spite of a larger statistics.
The main reason for this is that the COHERENT CsI detector measured less events than predicted
by the SM, while off-diagonal vector couplings can only increase the CE$\nu$NS cross section,
due to the absence of an interference term with the SM contribution. As a result, the CsI (2021) constraints
on these parameters are better than expected.
By contrast, flavor-diagonal vector couplings interfere with the SM contribution and
can therefore fit a deficit of observed CE$\nu$NS events.
This is the reason why, in Fig.~\ref{1D_proj_vector-new2},
the first minimum of the black dashed curves for flavor-diagonal vector couplings is shifted from zero.

\begin{figure}
\centering
\includegraphics[width=0.32\textwidth]{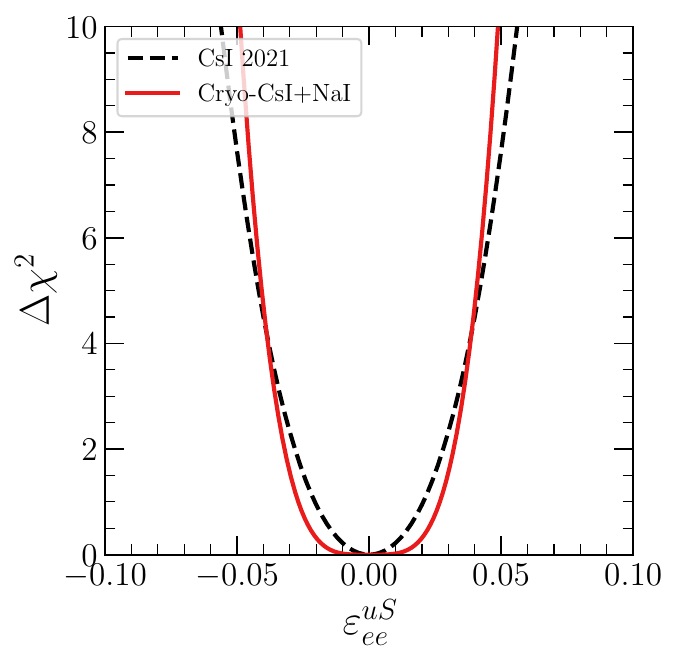}
\includegraphics[width=0.32\textwidth]{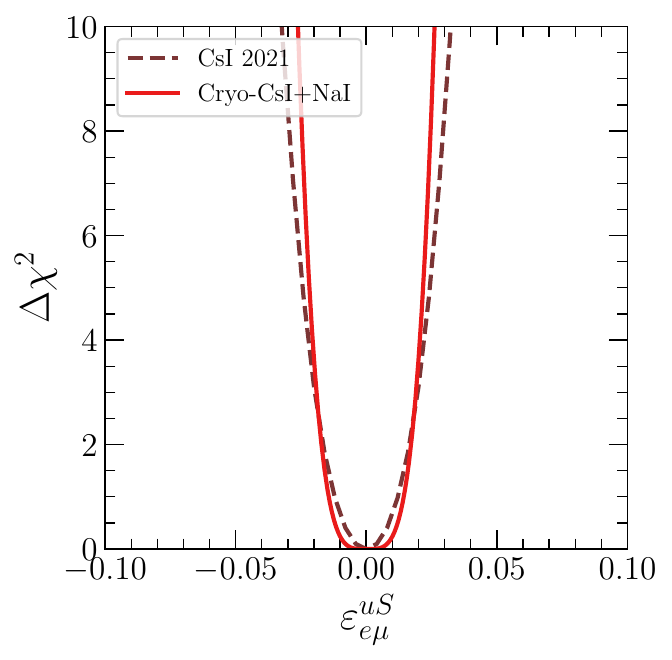}
\includegraphics[width=0.32\textwidth]{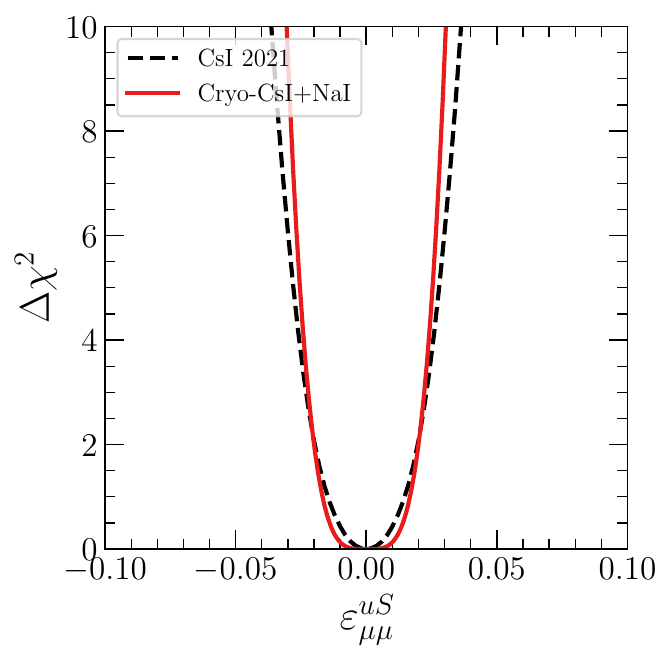}
\includegraphics[width=0.32\textwidth]{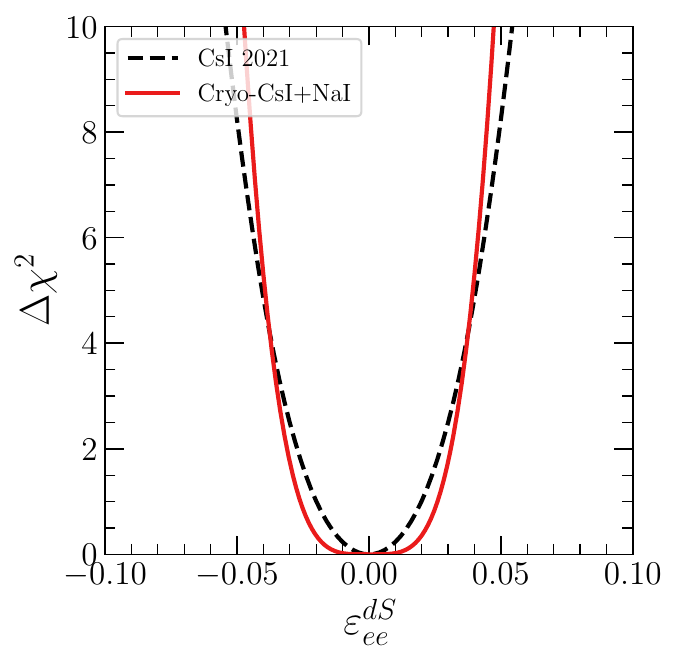}
\includegraphics[width=0.32\textwidth]{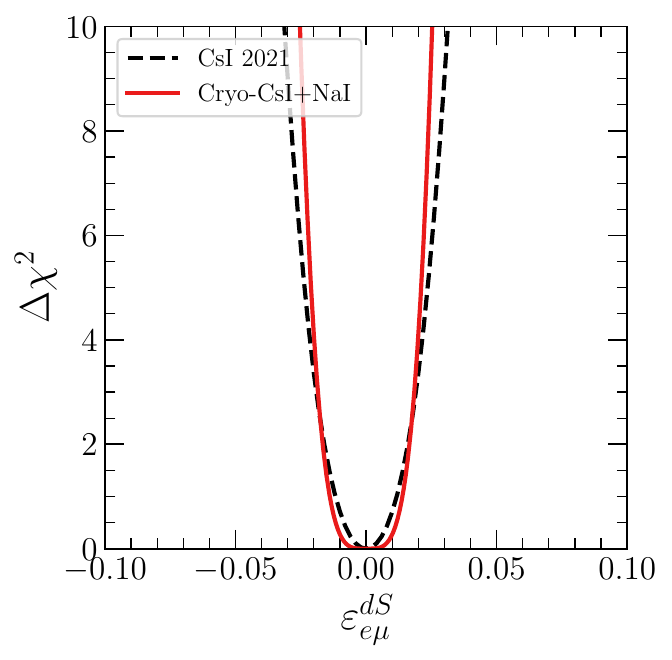}
\includegraphics[width=0.32\textwidth]{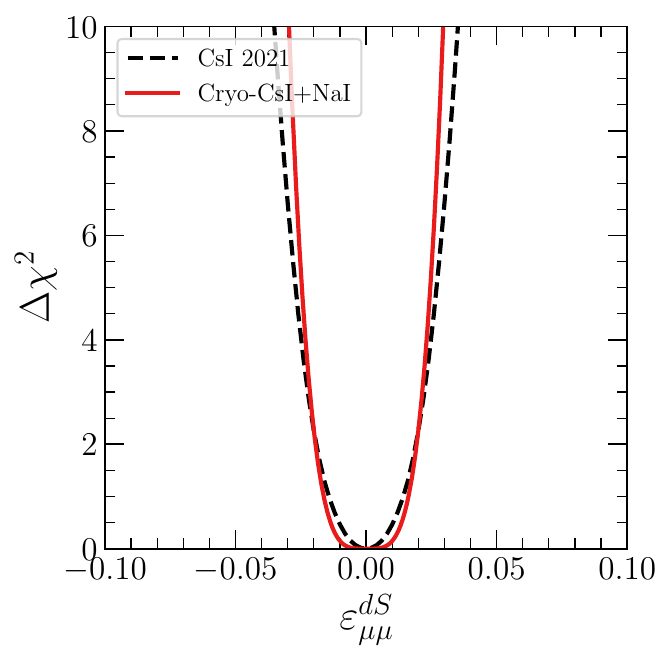}
\caption{Same as Fig.~\ref{1D_proj_vector-new2}, but for the scalar couplings $\varepsilon^{qS}_{ee}$ (left panels),
$\varepsilon^{qS}_{e\mu}$ (middle panels) and $\varepsilon^{qS}_{\mu\mu}$ (right panels),
with $q=u$ in the upper panels and $q=d$ in the lower panels.
All couplings are assumed to be real.}
\label{1D_proj_scalar}
\end{figure}

Let us now move on to the results for scalar GNIs. As explained before, CE$\nu$NS
measurements at the SNS only give six independent contraints on scalar couplings, since
$\varepsilon^{qS}_{e\tau}$ and $\varepsilon^{qS}_{\mu\tau}$ contribute the same way to the
CE$\nu$NS cross section as $\varepsilon^{qS}_{ee}$ and $\varepsilon^{qS}_{\mu\mu}$,
respectively\footnote{This statement holds for symmetric scalar couplings, hence for Majorana neutrinos
(as assumed in this paper), but not for Dirac neutrinos.}. We therefore only present results
for the 6 scalar couplings $\varepsilon^{qS}_{ee}$, $\varepsilon^{qS}_{\mu\mu}$
and $\varepsilon^{qS}_{e \mu}$ ($q = u, d$). The $\Delta \chi^2$ profiles are displayed
in Fig.~\ref{1D_proj_scalar}, with the same color code as before, and the $90\%$ C.L. and $2\sigma$
expected sensitivities of the ${\rm NaI} + {\rm Cryo}$-CsI detector combination,
together with the constraints from the COHERENT $\rm CsI$ 2021 data, are given in Table~\ref{const_table_scalar}.
As can be seen by comparing Figs.~\ref{1D_proj_vector-new2} and~\ref{1D_proj_scalar},
the results for scalar couplings are similar to the ones for off-diagonal vector couplings, which is not a surprise
since both types of interactions share the property of not interfering with the SM contribution in the CE$\nu$NS cross section.
In particular, we only expect a small improvement of the constraints on scalar couplings
from the ${\rm NaI} + {\rm Cryo}$-CsI detector combination, mainly visible at the
$\gtrsim 3\sigma$ confidence level.

\begin{table}
\newcommand{\mc}[3]{\multicolumn{#1}{#2}{#3}}
\newcommand{\mr}[3]{\multirow{#1}{#2}{#3}}
\centering
\begin{adjustbox}{width=1\textwidth}
\begin{tabular}{|c|c|c|c|c|c|c|c|}
\hline
 \mr{2}{*}{Detector} & \mr{2}{*}{$\hskip .2cm q \hskip .2cm$} & \mc{2}{c|}{$\varepsilon^{qS}_{ee}$} &  \mc{2}{c|}{$\varepsilon^{qS}_{e\mu}$} &  \mc{2}{c|}{$\varepsilon^{qS}_{\mu\mu}$}  \\ 
 \cline{3-4}
 \cline{5-6}
 \cline{7-8}
  &  & $90\%\,\rm C.L.$  & $2\sigma\,\rm C.L.$ & $90\%\,\rm C.L.$ & $2\sigma\,\rm C.L.$  &$90\%\,\rm C.L.$ & $2\sigma\,\rm C.L.$\\
\hline
\hline
 \mr{2}{*}{CsI (2021)}   &  $u$  & $\left[-0.032, 0.032\right]\,$  & $\left[-0.038, 0.038\right]\,  $ &  $\left[-0.018, 0.018\right]\,$ & $\left[-0.022, 0.022\right]\,$ &  $\left[-0.022, 0.022\right]\,$    & $\left[-0.026, 0.026\right]\,$     \\ 
  & $d$  & $\left[-0.031, 0.031\right]\, $ &$\left[-0.037, 0.037\right]\, $ &	  $\left[-0.018, 0.018\right]\,	$  &  $\left[-0.022, 0.022\right]\, 	$  & $\left[-0.022, 0.022\right]\,$ & $\left[-0.025, 0.025\right]\,$ \\
\hline
\mr{2}{*}{${\rm NaI} + {\rm Cryo}$-CsI}   &  $u$  & $\left[-0.034, 0.034\right]\,$  & $\left[-0.038, 0.038\right]\,  $ &  $\left[-0.018, 0.018\right]\,$     & $\left[-0.02, 0.02\right]\,$ & $\left[-0.021, 0.021\right]\,$ & $\left[-0.023, 0.023\right]\,$      \\ 
  & $d$  & $\left[-0.033, 0.033\right]\, $ &$\left[-0.037, 0.037\right]\, $ &	  $\left[-0.017, 0.017\right]\,	$  &  $\left[-0.019, 0.019\right]\, 	$  & $\left[-0.021, 0.021\right]\,$ & $\left[-0.023, 0.023\right]\,$  \\
\hline
\end{tabular}
\end{adjustbox}
\caption{Same as Table~\ref{const_table_vector}, but for the scalar couplings $\varepsilon^{qS}_{ee}$,
$\varepsilon^{qS}_{e\mu}$ and $\varepsilon^{qS}_{\mu\mu}$ ($q=u,\,d$).}
\label{const_table_scalar}
\end{table}

\begin{figure}
\centering
\includegraphics[width=0.32\textwidth]{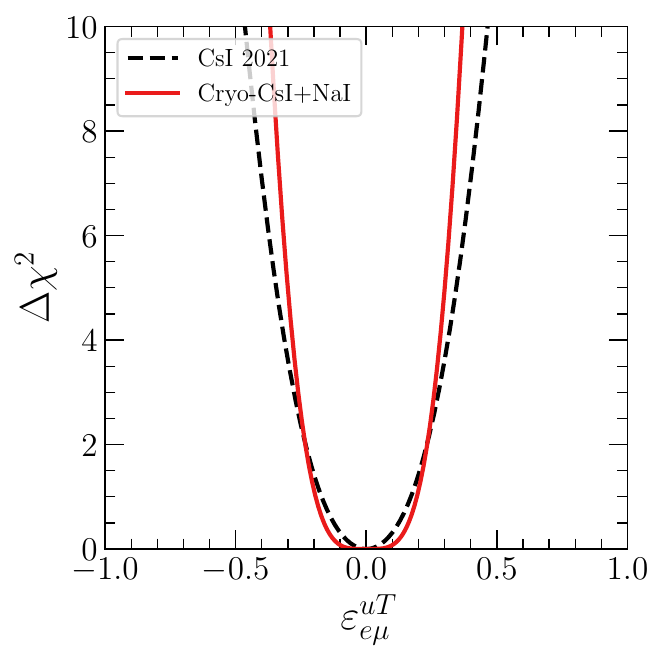}
\includegraphics[width=0.32\textwidth]{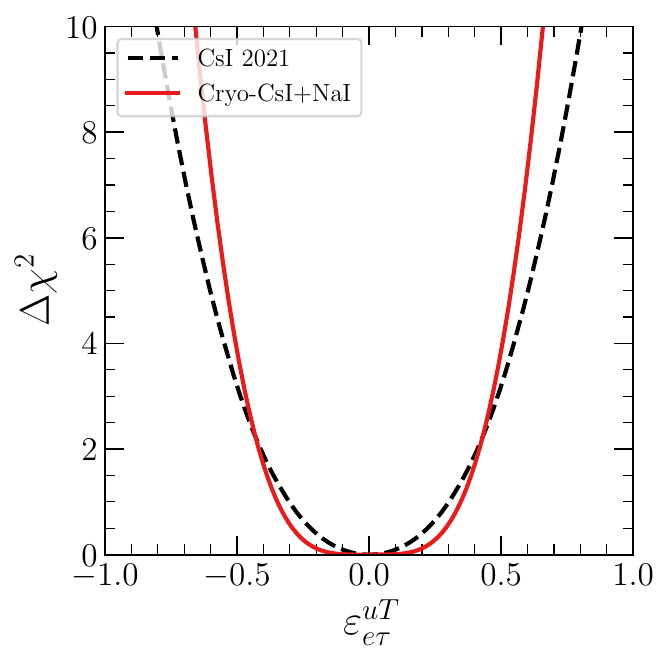}
\includegraphics[width=0.32\textwidth]{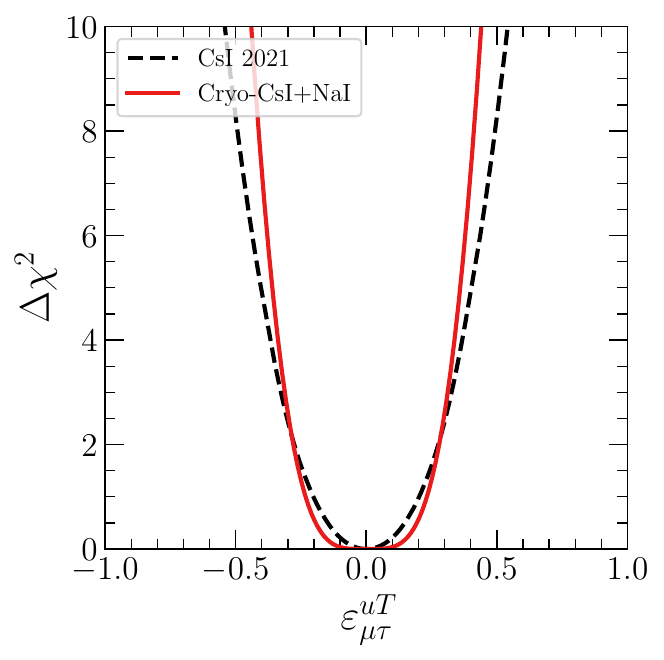}
\includegraphics[width=0.32\textwidth]{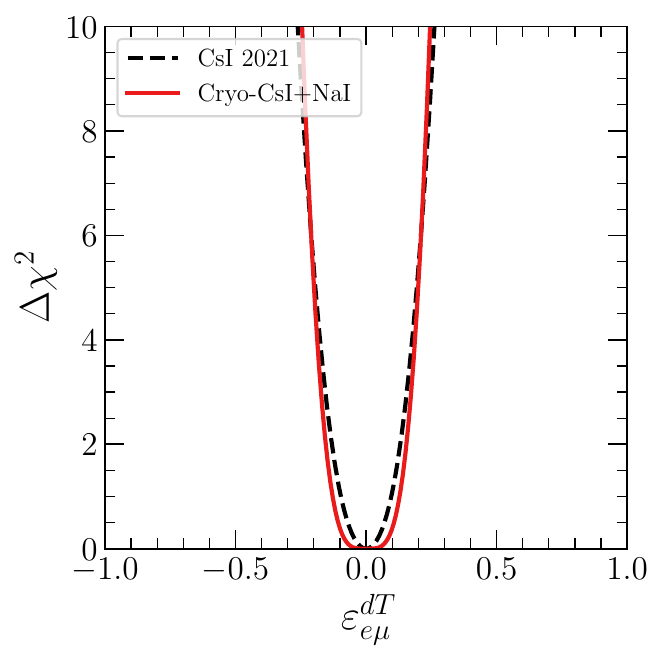}
\includegraphics[width=0.32\textwidth]{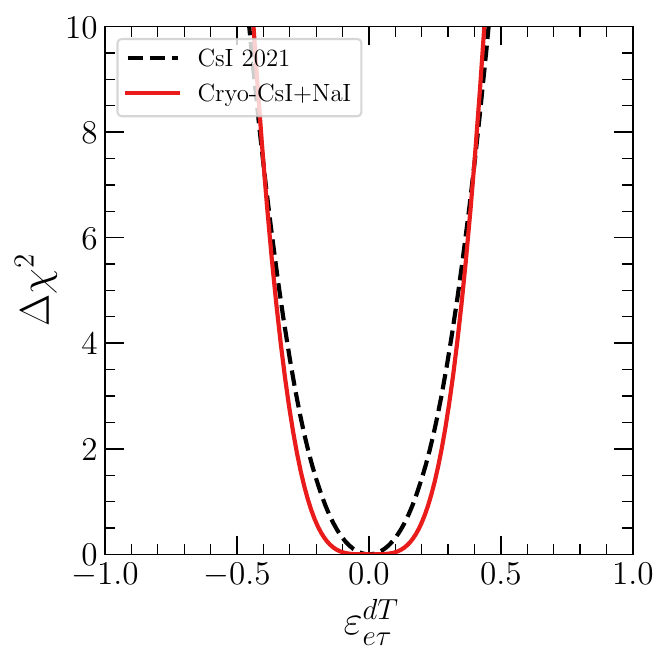}
\includegraphics[width=0.32\textwidth]{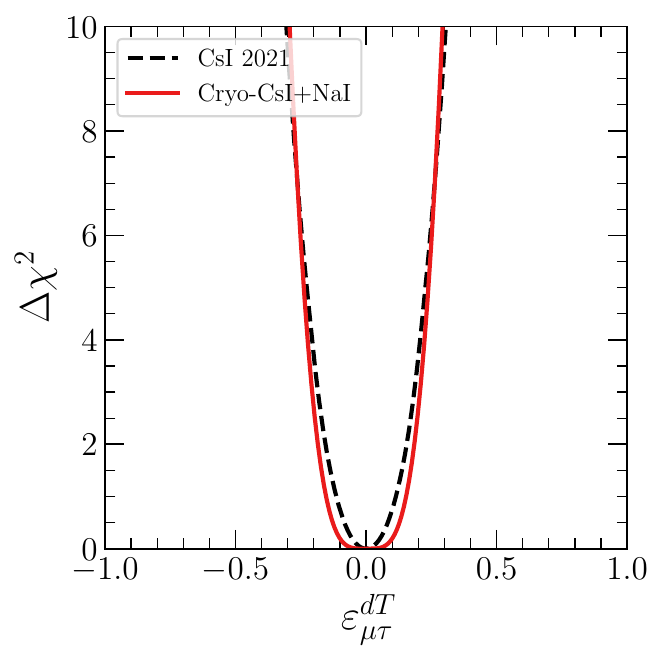}
\caption{Same as Fig.~\ref{1D_proj_vector-new2}, but for the tensor couplings $\varepsilon^{qT}_{e\mu}$ (left panels),
$\varepsilon^{qT}_{e\tau}$ (middle panels) and $\varepsilon^{qT}_{\mu\tau}$ (right panels),
with $q=u$ in the upper panels and $q=d$ in the lower panels.
All couplings are assumed to be real.}
\label{1D_proj_tensor}
\end{figure}

Finally, the results for tensor GNIs are shown in Fig.~\ref{1D_proj_tensor} and
Table~\ref{const_table_tensor}. Recall that there are only 6 independent tensor couplings
if, as assumed in this paper, neutrinos are Majorana fermions, namely $\varepsilon^{qT}_{e\mu}$,
$\varepsilon^{qT}_{e\tau}$ and $\varepsilon^{qT}_{\mu\tau}$ ($q = u, d$).
Comparing the expected sensitivities of the
${\rm NaI} + {\rm Cryo}$-CsI detector combination with the constraints from
the COHERENT CsI 2021 data (see Table~\ref{const_table_tensor}),
we only expect a small improvement of the constraints
on tensor coupligs involving an up quark.

\begin{table}
\newcommand{\mc}[3]{\multicolumn{#1}{#2}{#3}}
\newcommand{\mr}[3]{\multirow{#1}{#2}{#3}}
\centering
\begin{adjustbox}{width=1\textwidth}
\begin{tabular}{|c|c|c|c|c|c|c|c|}
\hline
 \mr{2}{*}{Detector} & \mr{2}{*}{$\hskip .2cm q \hskip .2cm$} & \mc{2}{c|}{$\varepsilon^{qT}_{e\mu}$} &  \mc{2}{c|}{$\varepsilon^{qT}_{e\tau}$} &  \mc{2}{c|}{$\varepsilon^{qT}_{\mu\tau}$}  \\ 
 \cline{3-4}
 \cline{5-6}
 \cline{7-8}
  &  & $90\%\,\rm C.L.$  & $2\sigma\,\rm C.L.$ & $90\%\,\rm C.L.$ & $2\sigma\,\rm C.L.$  &$90\%\,\rm C.L.$ & $2\sigma\,\rm C.L.$\\
\hline
\hline
 \mr{2}{*}{CsI (2021)}   &  $u$  &  $\left[-0.265, 0.265\right]\,$     & $\left[-0.314, 0.314\right]\,$ & $\left[-0.467, 0.467\right]\,$  & $\left[-0.55, 0.55\right]\, $ &      $\left[-0.314, 0.314\right]\,$  &  $\left[-0.369, 0.369\right]\,$    \\
  & $d$  &	  $\left[-0.15, 0.15\right]\,	$  &  $\left[-0.177, 0.177\right]\, 	$ & $\left[-0.264, 0.264\right]\, $ &$\left[-0.310, 0.310\right]\, $  & $\left[-0.177, 0.177\right]\,$ & $\left[-0.208, 0.208\right]\,$ \\
\hline
\mr{2}{*}{${\rm NaI} + {\rm Cryo}$-CsI}   &  $u$  &  $\left[-0.253, 0.253\right]\,$  & $\left[-0.282, 0.282\right]\,$ & $\left[-0.453, 0.453\right]\,$  & $\left[-0.505, 0.505\right]\,  $ &   $\left[-0.303, 0.303\right]\,$  &  $\left[-0.337, 0.337\right]\,$    \\
  & $d$  &	  $\left[-0.167, 0.167\right]\,	$  &  $\left[-0.187, 0.187\right]\, 	$ & $\left[-0.299, 0.299\right]\, $ &$\left[-0.334, 0.334\right]\, $  & $\left[-0.200, 0.200\right]\,$ &  $\left[-0.223, 0.223\right]\,$\\  
\hline
\end{tabular}
\end{adjustbox}
\caption{Same as Table~\ref{const_table_vector}, but for the tensor couplings $\varepsilon^{qT}_{e\mu}$,
$\varepsilon^{qT}_{e\tau}$ and $\varepsilon^{qT}_{\mu\tau}$ ($q=u,\,d$).}
\label{const_table_tensor}
\end{table}

\subsection{Expected sensitivities to pairs of GNI parameters}
\label{subsec:two_GNIs}

\begin{figure}
\centering
\includegraphics[width=0.49\textwidth]{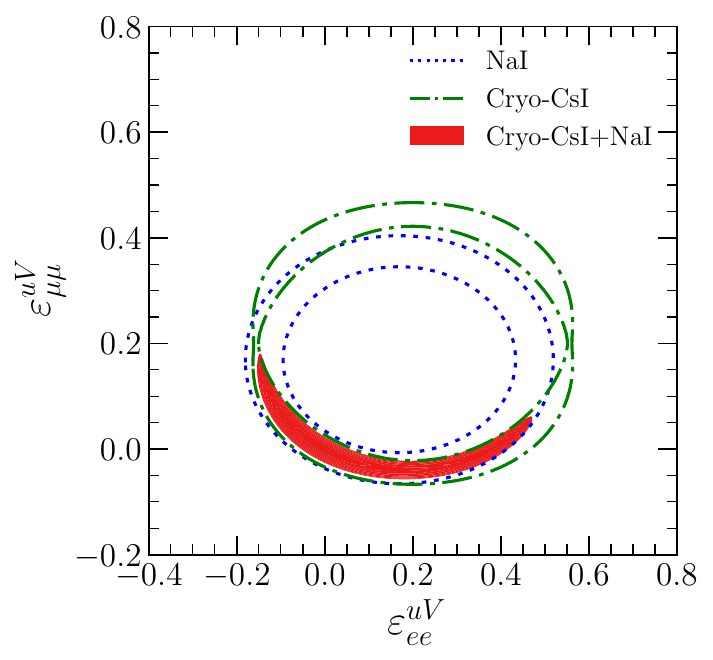}
\includegraphics[width=0.49\textwidth]{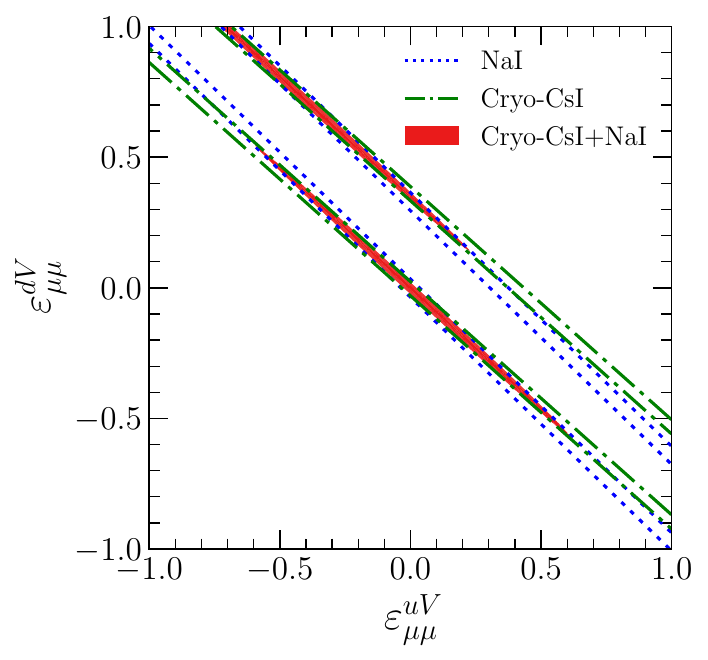}
\caption{Expected 90\% C.L. allowed regions (with two degrees of freedom, i.e. $\Delta \chi^2 \leq 4.61$)
for the pairs of NSI parameters $(\varepsilon^{uV}_{ee},\, \varepsilon^{uV}_{\mu\mu})$ (left panel)
and $(\varepsilon^{uV}_{\mu\mu},\, \varepsilon^{dV}_{\mu\mu})$ (right panel).
The blue dotted contours (resp. the green dash-dotted contours) correspond to the expected constraints
from the NaI detector described in Subsection~\ref{subsec:NaI_detector}
(resp. the Cryo-CsI detector described in Subsection~\ref{subsec:Cryo-CsI_detector}),
assuming 3 years of data taking at the SNS.
The red regions represent the expected combined sensitivities of the two detectors.}
\label{contour_vector-new}
\end{figure}

In this section, we study the expected sensitivity of the future NaI and Cryo-CsI detectors to pairs of GNI parameters,
i.e. we consider two nonvanishing couplings at a time. As is well known,
this situation leads to degeneracies in the form of extended allowed regions in the two-dimensional
GNI parameter space.
Combining data on target nuclei characterized by different proton to neutron ratios can help to reduce these
degeneracies, as shown e.g. in Refs.~\cite{Barranco:2005yy,Baxter:2019mcx,Chatterjee:2022mmu}.
In the case of the future NaI and cryogenic CsI detectors studied in this paper, the target materials
significantly differ by their proton to neutron ratios: $0.92$ for sodium (the NaI detector
is not sensitive to iodine recoils) versus $0.71$ and $0.72$ for cesium and iodine,
both of which contribute to the CE$\nu$NS rate in the Cryo-CsI detector.

We begin our analysis by studying the case where the two nonvanishing GNI parameters are of vector type
(i.e., NSIs). Many different pairs of NSI parameters can be considered;
we only present results for a few representative examples, as the other possible choices are characterized
by similar allowed regions. As we did in the previous subsection, we start our discussion
by presenting in Fig.~\ref{contour_vector-new} the expected individual sensitivities of the future NaI and Cryo-CsI detectors
to two different pairs of (flavor-diagonal) NSI parameters:
$(\varepsilon^{uV}_{ee},\, \varepsilon^{uV}_{\mu\mu})$ in the left panel,
and $(\varepsilon^{uV}_{\mu\mu},\, \varepsilon^{dV}_{\mu\mu})$ in the right panel.
The expected 90\% C.L. allowed regions are enclosed by blue dotted lines for NaI
and by green dash-dotted contours for Cryo-CsI, assuming as in the previous subsection
3 years of data taking and $\sigma_{sig} = 10\%$, $ \sigma_{bg} = 5\%$.

The elliptic-like shape of the allowed regions
for the NSI parameters $(\varepsilon_{ee}^{uV}, \varepsilon_{\mu\mu}^{uV})$
(left panel of Fig.~\ref{contour_vector-new})
can be understood from the expression for the predicted CE$\nu$NS rate in each reconstructed nuclear
recoil energy bin~\cite{Chatterjee:2022mmu}.
For a single-nucleus detector\footnote{The NaI detector is only sensitive to Na recoils, so can
be considered single-nucleus. As for the Cryo-CsI detector, its two target nuclei have approximately
the same numbers of protons and neutrons.},
the predicted number of events in the recoil energy bin $[T_i, T_{i+1}]$ can be written as
\begin{equation}
  N_i^{th}\, =\, C_i^e \left ( 2Z + N \right )^2 \left ( \varepsilon^{uV}_{ee} + \frac{Zg_{V}^p + Ng_{V}^n}{2Z + N} \right )^2 
    +\, C_i^\mu \left ( 2Z + N \right )^2 \left ( \varepsilon^{uV}_{\mu\mu} + \frac{Zg_{V}^p + Ng_{V}^n}{2Z + N} \right )^2\, ,
\label{eq:ellipse}
\end{equation}
where $C^e_i$ (resp. $C^\mu_i$) is the number obtained by performing the integral over $E_{\nu_e}$
(resp. $E_{\nu_\mu}$ and $E_{\bar \nu_\mu}$) and the true and reconstructed nuclear recoil energies
in Eq.~(\ref{eq:CEvNS_events_NR_bins}), after having factorized out $|C^V_{\alpha \alpha} + Q_{W, \alpha}|^2$.
Eq.~(\ref{eq:ellipse}) is the equation of an ellipse with semi-major and semi-minor axes (given respectively
by $\sqrt{N_i^{th}/C_i^e \left( 2Z + N \right)^2}$ and $\sqrt{N_i^{th}/C_i^\mu \left( 2Z + N \right)^2}\,$)
that depend on both the nucleus and the energy bin (the center of the ellipse also depends
on the nucleus). All sets of ($\varepsilon^{uV}_{ee}$,~$\varepsilon^{uV}_{\mu \mu}$)
values lying on this ellipse predict the same number of CE$\nu$NS events in the $i$th energy bin, but not
necessarily in the other bins, resulting in an approximate degeneracy between these two parameters.
This explains why the expected 90\% C.L. region allowed by the Cryo-CsI detector
has the form of an approximate, partly broken ellipse (for the NaI detector,
the allowed region turns out to be an almost perfect ellipse).
Note that the NaI and Cryo-CsI allowed areas do not fully overlap, as both the center
and the semi-axes of the ellipses depend on the nucleus.
If we now consider the  90\% C.L. region allowed
by the combination of the future NaI and Cryo-CsI detectors, shown in red in the left panel of Fig.~\ref{contour_vector-new},
we can see that the degeneracy between the vector couplings $\varepsilon_{ee}^{uV}$ and $\varepsilon_{\mu\mu}^{uV}$
is partially broken. Namely, the allowed region reduces to a crescent, approximately corresponding
to the intersection of the NaI and Cryo-CsI ellipses. This result illustrates the benefit of combining
data from detectors with different proton-to-neutron ratios in order to reduce degeneracies
between flavor-diagonal NSI parameters.

Let us now consider the right panel of Fig.~\ref{contour_vector-new}, which displays the 90\% C.L.
allowed regions for the flavor-diagonal NSI parameters $(\varepsilon_{\mu\mu}^{uV}, \varepsilon_{\mu\mu}^{dV})$.
In this case, the individual NaI and Cryo-CsI allowed regions consist of two parallel bands~\cite{Scholberg:2005qs}.
This is a consequence of the fact that, for a single-nucleus target (or a target composed of two nuclei
with approximately the same proton to neutron ratio, like CsI), the CE$\nu$NS cross section
depends on the flavor-diagonal NSI parameters $\varepsilon^{uV}_{\alpha \alpha}$ and $\varepsilon^{dV}_{\alpha \alpha}$
only through the modulus of the combination
\begin{equation}
  C^V_{\alpha \alpha} + Q_{W, \alpha}\, \simeq\, (Z + 2 N)
    \left( \varepsilon^{dV}_{\alpha \alpha} - m\, \varepsilon^{uV}_{\alpha \alpha} \right) - \frac{1}{2}\ ,
    \qquad m = -\frac{2Z+N}{Z+2N}\ ,
\label{eq:weak_charge_NSIs_bis}
\end{equation}
in which we neglected a term suppressed by $g^p_V \ll 1$.
Thus, values of $\varepsilon^{uV}_{\alpha \alpha}$ and $\varepsilon^{dV}_{\alpha \alpha}$ lying
on a straight ligne with slope $m$~\cite{Barranco:2005yy} produce the same number of CE$\nu$NS
events in each recoil energy bin.
This is an example of the degeneracies mentioned above.
It follows that the parameter space region allowed by SM-like experimental data at a given confidence interval
(here 90\% C.L.) consists of two parallel bands around the straight lines
$\varepsilon^{dV}_{\alpha \alpha} - m\, \varepsilon^{uV}_{\alpha \alpha} = 0$
and $\varepsilon^{dV}_{\alpha \alpha} - m\, \varepsilon^{uV}_{\alpha \alpha} = 1 / (Z + 2 N)$.
This is indeed what we can observe in the right panel of Fig.~\ref{contour_vector-new}
for the individual NaI and Cryo-CsI allowed regions.
Note that the slopes of the bands differ
due to the different proton-to-neutron ratios of the two detectors. Thus, combining the results
of the two detectors leads to a partial breaking of the degeneracies between the flavor-diagonal
NSI parameters $\varepsilon_{\mu\mu}^{uV}$ and $\varepsilon_{\mu\mu}^{dV}$,
as can be seen in the right panel of Fig.~\ref{contour_vector-new}, where the two infinite bands reduce
to the red closed areas).

\begin{figure}
\centering
\includegraphics[width=0.48\textwidth]{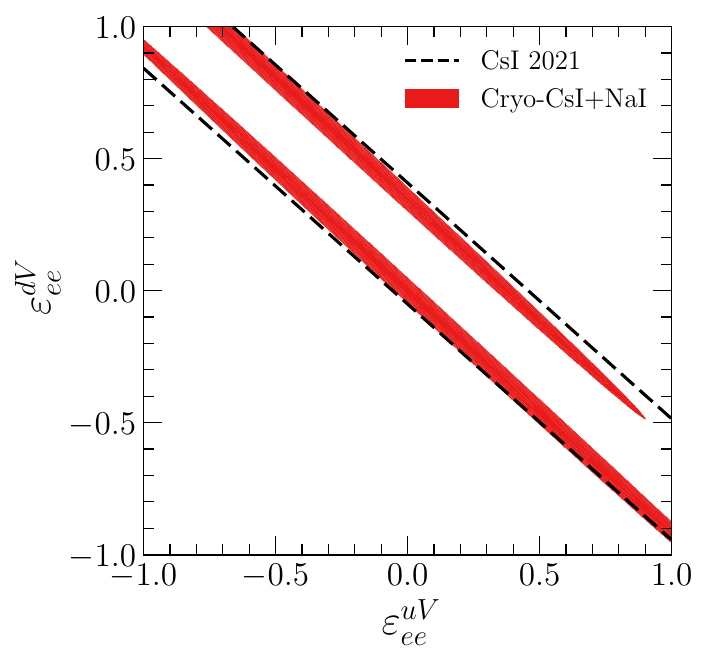}
\includegraphics[width=0.48\textwidth]{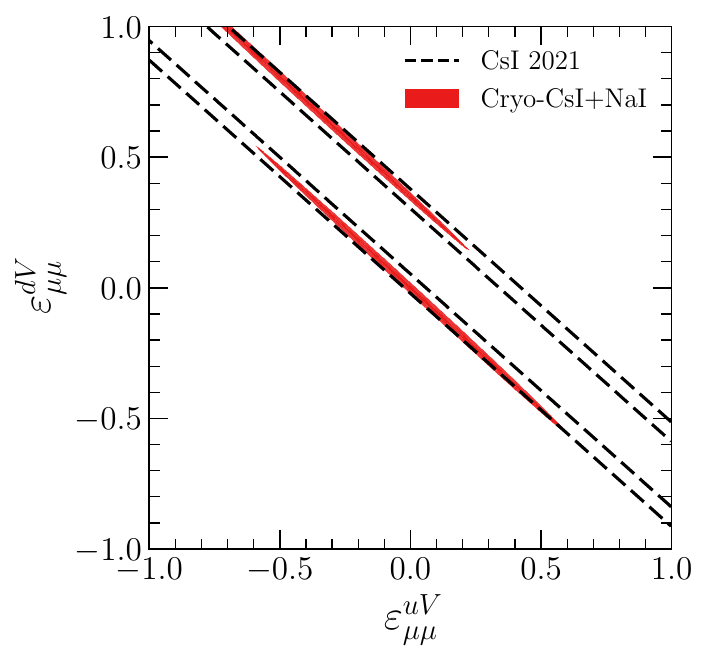}
\includegraphics[width=0.48\textwidth]{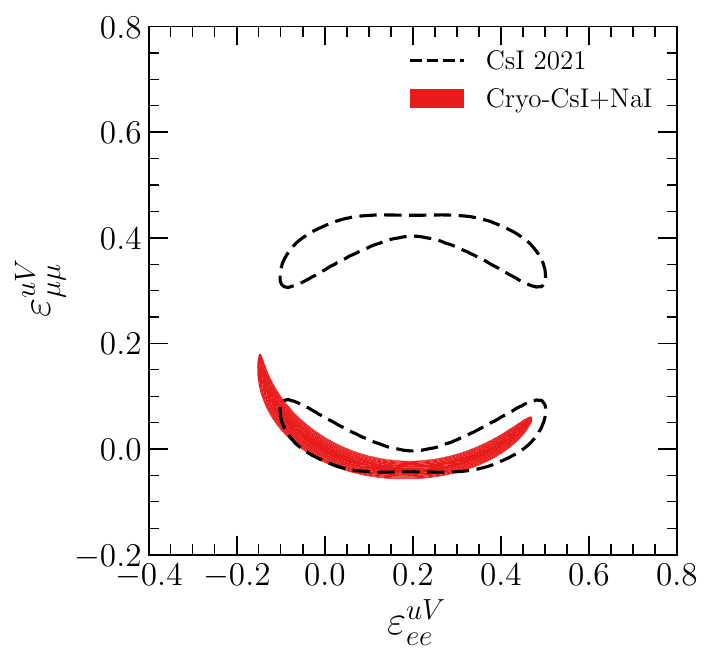}
\includegraphics[width=0.48\textwidth]{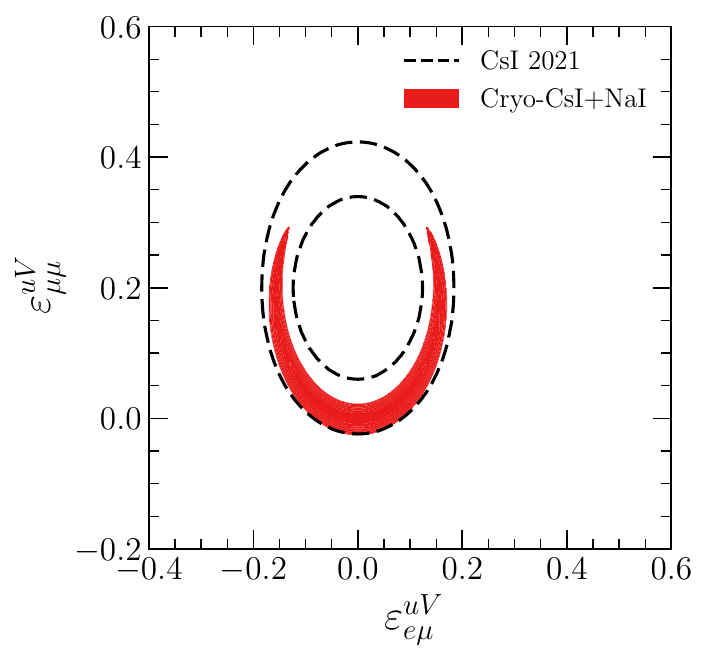}
\caption{90\% C.L. allowed regions (with two degrees of freedom, i.e. $\Delta \chi^2 \leq 4.61$)
for different pairs of NSI parameters:
$(\varepsilon^{uV}_{ee},\, \varepsilon^{dV}_{ee})$ (upper left panel),
$(\varepsilon^{uV}_{\mu\mu},\, \varepsilon^{dV}_{\mu\mu})$ (upper right panel),
$(\varepsilon^{uV}_{ee},\, \varepsilon^{uV}_{\mu\mu})$ (lower left panel)
and $(\varepsilon^{uV}_{e\mu},\, \varepsilon^{uV}_{\mu\mu})$ (lower right panel).
The regions enclosed by black dashed lines correspond to the constraints
from the currently available CsI data, while the red areas
represent the expected sensitivities of the combined ${\rm NaI} + {\rm Cryo}$-CsI analysis
with the detector characteristics described in Section~\ref{sec:experiment},
assuming 3 years of data taking at the SNS for each detector.
The flavor off-diagonal parameter $\varepsilon^{qV}_{e\mu}$ is assumed to be real.}
\label{2D_proj_vector}
\end{figure}

Now that we have shown the impact of a combined ${\rm NaI} + {\rm Cryo}$-CsI analysis
on degeneracies between flavor-diagonal NSI parameters,
we are ready to compare the expected sensitivities of this combined analysis with the constraints
obtained from the COHERENT CsI 2021 data.
Fig.~\ref{2D_proj_vector} displays the allowed regions at the 90\% confidence level for four different pairs
of vector couplings. The upper panels correspond to combinations of flavor-diagonal couplings involving the same lepton flavor
but different quarks, namely $\left(\varepsilon_{ee}^{uV},\, \varepsilon_{ee}^{dV}\right)$ in the left panel
and $\left(\varepsilon_{\mu\mu}^{uV},\, \varepsilon_{\mu\mu}^{dV}\right)$ in the right panel.
By contrast, the lower panels correspond to pairs of vector couplings involving the same quark
but different lepton flavors. The combination in the left panel, $\left(\varepsilon_{ee}^{uV},\, \varepsilon_{\mu\mu}^{uV}\right)$,
involves two flavor-diagonal couplings, while the one in the right panel, $\left(\varepsilon_{e\mu}^{uV},\, \varepsilon_{\mu\mu}^{uV}\right)$,
involves a flavor-diagonal and an off-diagonal coupling.
In all cases, the regions enclosed by black dashed  lines are the ones allowed
at the 90\% confidence level by the current CsI data (these allowed regions were first presenteded
in Ref.~\cite{DeRomeri:2022twg}, and are shown here for reference), while the red areas 
correspond to the expected sensitivity of the ${\rm NaI} + {\rm Cryo}$-CsI combination,
assuming 3 years of data taking at the SNS for each detector.
For all pairs of NSI parameters considered here, the expected future sensitivities
represent a clear improvement on the current CsI constraints.
This is due to two main factors: the increase in statistics, and the different proton-to-neutron ratios
of the NaI and CsI targets, whose impact on parameter degeneracies has been discussed above.

In the upper panels, in addition to the degeneracy breaking, the effect of the larger statistics
is clearly visible, especially in the  $\left(\varepsilon_{ee}^{uV},\, \varepsilon_{ee}^{dV}\right)$ parameter space.
While the two bands associated with the current CsI detector overlap, due to the limited statistics
in the $\nu_e$ channel (there are twice as many $\nu_\mu$'s and $\overline{\nu}_\mu$'s as
$\nu_e$'s in the SNS neutrino beam), the expected allowed region for the
${\rm NaI} + {\rm Cryo}$-CsI combination consists of two disconnected areas.
For the pairs of NSI parameters $\left(\varepsilon_{ee}^{uV},\, \varepsilon_{\mu\mu}^{uV}\right)$ (lower left panel),
the region allowed by the current CsI detector is a broken ellipse, with no points around
$\varepsilon_{\mu\mu}^{uV} \simeq 1 / 2 (2 Z/N + 1) \approx 0.2$, where the NSI and SM
contributions to $\nu_\mu$- and $\overline \nu_\mu$-induced CE$\nu$NS completely cancel out.
This leaves two degenerate allowed regions around $\varepsilon_{\mu\mu}^{uV} = 0$ and
$\varepsilon_{\mu\mu}^{uV} \simeq 1 / (2 Z/N + 1) \approx 0.4$, while only the red area
is expected to survive the degeneracy breaking enforced by the ${\rm NaI} + {\rm Cryo}$-CsI combination
(see the discussion around Eq.~(\ref{eq:ellipse})).

Finally, we also observe an ellipse-shaped degeneracy between the vector couplings
$\varepsilon_{e\mu}^{uV}$ and $\varepsilon_{\mu\mu}^{uV}$ in the lower right panel of Fig.~\ref{2D_proj_vector},
which is not broken by the current CsI data. This can be understood by noting that $\varepsilon_{e\mu}^{uV}$
induces both $\nu_e$ and $\nu_\mu$/$\overline \nu_\mu$ coherent scatterings on the target nuclei,
without interfering with the SM contribution. Nonvanishing values of $\varepsilon_{e\mu}^{uV}$
can therefore compensate for the destructive interference between the SM and $\varepsilon_{\mu\mu}^{uV}$
contributions to the CE$\nu$NS cross section, which would otherwise reduce the number
of $\nu_\mu$/$\overline \nu_\mu$-induced signal events. As a result, the ellipse is not broken
around $\varepsilon_{\mu \mu}^{uV} \approx 0.2$ in the case of the current CsI detector.
This degeneracy is, however, expected to be broken by the ${\rm NaI} + {\rm Cryo}$-CsI
combination, for the same reason as for the $\left(\varepsilon_{ee}^{uV},\, \varepsilon_{\mu\mu}^{uV}\right)$
pair: due to the different proton-to-neutron ratios of the NaI and CsI targets, the corresponding ellipses
do not fully overlap, leaving only the allowed red region in the lower right panel of Fig.~\ref{2D_proj_vector}.

 \begin{figure}
\centering
\includegraphics[width=0.49\textwidth]{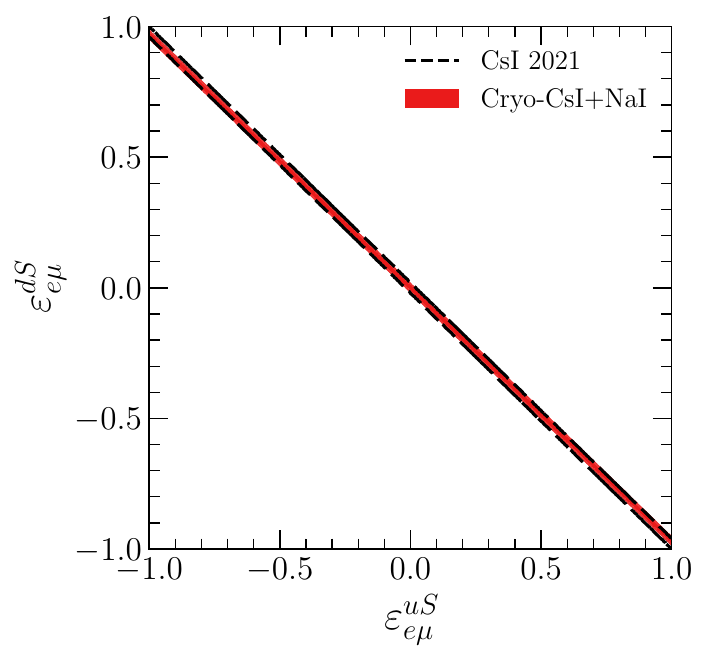}
\includegraphics[width=0.49\textwidth]{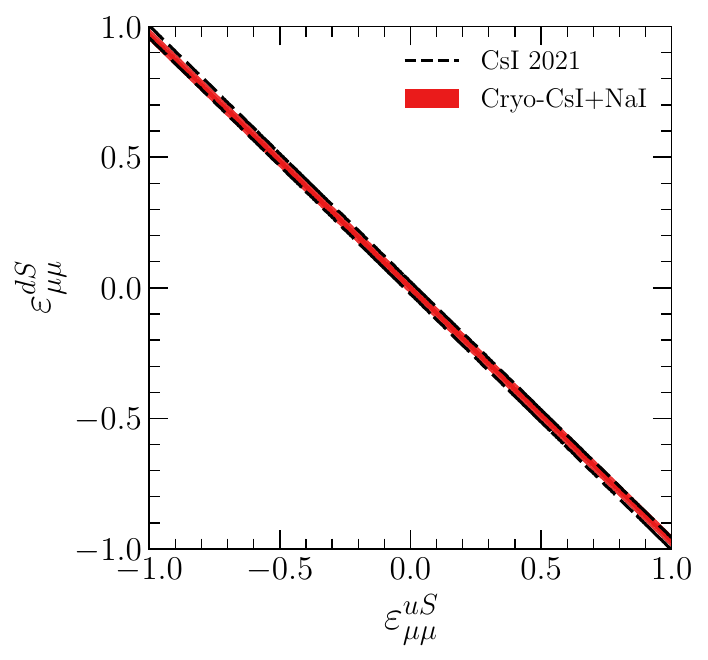}\\
\caption{90\% C.L. allowed regions for the pairs of scalar GNI parameters
$\left(\varepsilon_{e\mu}^{uS},\, \varepsilon_{e\mu}^{dS}\right)$ (left panel)
 and $\left(\varepsilon_{\mu\mu}^{uS},\, \varepsilon_{\mu\mu}^{dS}\right)$ (right panel).
The regions enclosed by black dashed lines correspond to the constraints
from the currently available CsI data, while the red areas
represent the expected sensitivities of the combined ${\rm NaI} + {\rm Cryo}$-CsI analysis
with the detector characteristics described in Section~\ref{sec:experiment},
assuming 3 years of data taking at the SNS for each detector.
All GNI parameters are assumed to be real.}
\label{2D_proj_scalar}
\end{figure}

\begin{figure}
\centering
\includegraphics[width=0.49\textwidth]{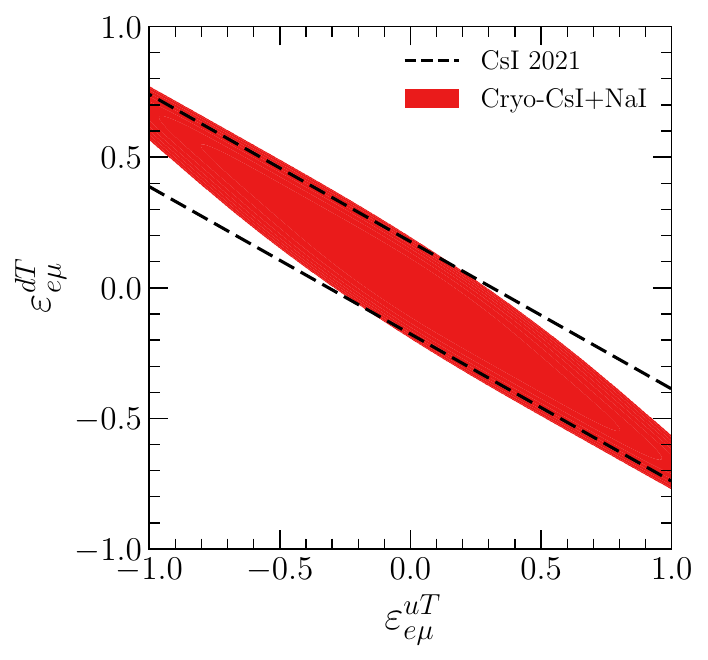}
\includegraphics[width=0.49\textwidth]{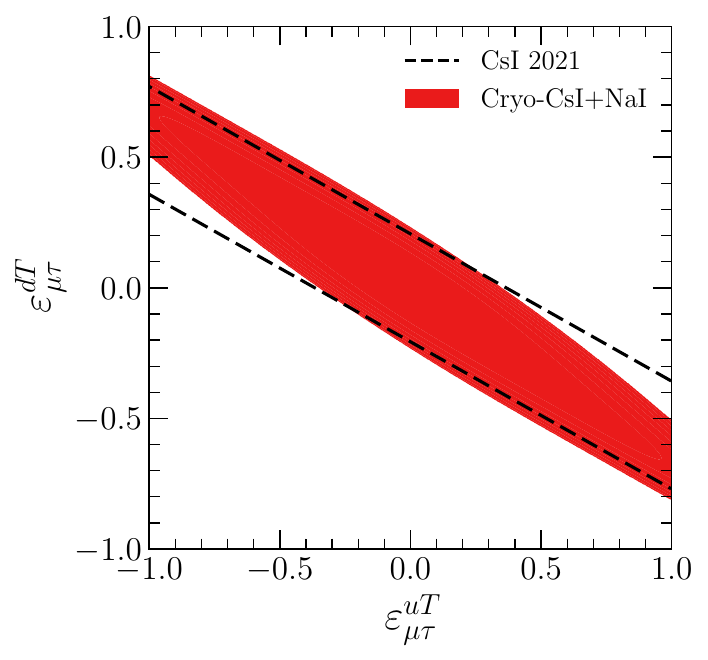}
\caption{Same as Fig.~\ref{2D_proj_scalar}, but for the pairs of tensor GNI parameters
$\left(\varepsilon_{e\mu}^{uT},\, \varepsilon_{e\mu}^{dT}\right)$ (left panel)
and $\left(\varepsilon_{\mu\tau}^{uT},\, \varepsilon_{\mu\tau}^{dT}\right)$ (right panel).
All GNI parameters are assumed to be real.}
\label{2D_proj_tensor}
\end{figure}

Let us now consider scalar and tensor GNIs. Fig.~\ref{2D_proj_scalar} shows the 90\% C.L.
allowed regions for two different pairs of scalar couplings with the same lepton flavor indices:
$\left(\varepsilon_{e\mu}^{uS},\, \varepsilon_{e\mu}^{dS}\right)$ in the left panel,
and $\left(\varepsilon_{\mu\mu}^{uS},\, \varepsilon_{\mu\mu}^{dS}\right)$ in the right panel.
As in Fig.~\ref{2D_proj_vector}, the regions allowed by the current CsI data are within black dashed lines,
while the red areas correspond to the expected sensitivity of the ${\rm NaI} + {\rm Cryo}$-CsI combination,
assuming 3 years of data taking at the SNS for each detector.
Since the scalar couplings $\varepsilon_{\alpha \beta}^{uS}$ and $\varepsilon_{\alpha \beta}^{dS}$
contribute to the CE$\nu$NS cross section through the linear combination $C^S_{\alpha \beta}$,
given in Eq.~(\ref{eq:scalar}), the allowed regions are straight bands around a line passing through
the origin (as a consequence of the fact that scalar couplings do not interfere with the SM contribution).
While the slope of this line
depends on the proton to neutron ratio of the target nucleus, it is in practice very close to $-1$ for Cs, I and Na.
A consequence of this is that even though sodium has a different proton to neutron ratio
than cesium and iodine, the ${\rm NaI} + {\rm Cryo}$-CsI combination is unable to reduce
the degeneracy between $\varepsilon_{\alpha \beta}^{uS}$ and $\varepsilon_{\alpha \beta}^{dS}$.

Finally, Fig.~\ref{2D_proj_tensor} displays the 90\% C.L. allowed regions for the pairs of tensor couplings
$\left(\varepsilon_{e\mu}^{uT},\, \varepsilon_{e\mu}^{dT}\right)$ (left panel)
and $\left(\varepsilon_{\mu\tau}^{uT},\, \varepsilon_{\mu\tau}^{dT}\right)$ (right panel).
Qualitatively similar features to the scalar case can be observed,
except that CE$\nu$NS is much less sensitive to tensor couplings,
and that the line slope has a different dependence on the proton to neutron ratio of the target nucleus.
This explains the shape of the expected allowed ${\rm NaI} + {\rm Cryo}$-CsI region,
but the poor sensitivity of CE$\nu$NS to tensor GNI parameters makes it difficult to reduce
the degeneracy between $\varepsilon_{\alpha \beta}^{uT}$ and $\varepsilon_{\alpha \beta}^{dT}$
more significantly than can be seen in Fig.~\ref{2D_proj_tensor}.

\section{Conclusions}
\label{sec:conclusions}

In this work, we studied the potential of two future detectors to constrain
non-standard and generalized neutrino interactions
via CE$\nu$NS measurements at the Spallation Neutron Source.
We considered a NaI detector with characteristics similar to the one that is currently being deployed
by the COHERENT collaboration, and a recently proposed cryogenic CsI detector.
These detectors have the interesting property that
their target materials are composed of nuclei with significantly different proton to neutron ratios,
a feature that is known to be useful for breaking degeneracies among NSI parameters.
In addition, they will benefit from a larger statistics than the current CsI detector.
We showed that these properties
of the future NaI and Cryo-CsI detectors make a combined analysis of their data particularly efficient
at breaking degeneracies involving flavor-diagonal NSI parameters, thus strongly improving
the current COHERENT constraints on non-standard neutrino interactions.
By contrast, no significant improvement is expected for neutrino interactions
whose contribution to the CE$\nu$NS cross section does not interfere with the SM one
(flavor off-diagonal NSIs, scalar and tensor generalized neutrino interactions),
unless an important statistical increase is achieved.

\subsection*{ACKNOWLEDGMENTS}
The work of S.S.C. is funded by the Deutsche Forschungsgemeinschaft (DFG, German Research Foundation) – project number 510963981. S. S. C. also acknowledges financial support from the LabEx P2IO (ANR-10-LABX-0038 - Project ``BSMNu'') in the framework of the ``Investissements d'Avenir'' (ANR-11-IDEX-0003-01 ) managed by the Agence Nationale de la Recherche (ANR), France.  This work has been supported by the Spanish grants PID2020-113775GB-I00 (MCIN/AEI/ 10.13039/501100011033) and CIPROM/2021/054 (Generalitat Valenciana). G.S.G. acknowledges financial support by the CIAPOS/2022/254 grant funded by
Generalitat Valenciana.
The work of SL is supported in part by the European Union's Horizon 2020 research and innovation programme under the Marie Sklodowska-Curie grant agreement No. 860881-HIDDeN.
O. G. M. was supported by the CONAHCyT grant 23238 and by SNII-Mexico.

\vfill
\eject

\bibliographystyle{utphys}
\bibliography{coherent}

\end{document}